# Microtransit revenue management informed by citywide travel demand and joint subscription–mode choice modeling

Xiyuan Ren[1*], Joseph Y. J. Chow[1], Venktesh Pandey[2], Linfei Yuan[1]

*1. C2SMARTER University Transportation Center, New York University Tandon School of Engineering, Brooklyn, USA*

*2. Department of Civil, Architectural, and Environmental Engineering*

*North Carolina Agricultural and Technical State University College of Engineering, Greensboro, USA*

* Corresponding author: xr2006@nyu.edu

## Abstract

As an IT-enabled multi-passenger mobility service, microtransit can improve accessibility, reduce congestion, and promote sustainability. However, realizing its business potential requires a deeper understanding of traveler preferences, highlighting the need for more effective tools for demand forecasting and revenue management, especially when actual usage data are limited. We propose an innovative modeling approach that integrates travel behavioral insights into microtransit policymaking. The approach operates by (1) leveraging citywide synthetic data to achieve greater spatiotemporal granularity, (2) estimating a nonparametric nested model for joint travel mode and ride-pass subscription choices, and (3) employing a simulation-based method to calculate revenue and traveler benefits under various policy scenarios. We demonstrate the applicability of our approach through a case study in Arlington, TX, one of the largest deployments of microtransit (Via) in the U.S. Using the simulation-based workflow, we evaluate alternative policy scenarios, including ride-pass discounts, event-based subsidies, and place-based subsidies, to assess their impacts on microtransit ridership, system revenue, and traveler welfare. The results indicate that reducing the weekly pass price from \$25 to \$18.9 and the monthly pass price from \$80 to \$71.5 would increase total revenue by approximately \$127 per day. A 100% trip fare discount could reduce 61 car trips to AT&T Stadium during a game event while generating an additional 82 microtransit trips per day to Medical City Arlington. However, achieving these mode shifts would require subsidies of approximately \$533 per event and \$483 per day, respectively. A sensitivity analysis reveals that the framework's predictions remain robust at the current adoption level, but accuracy becomes increasingly consequential as market share grows. The findings offer practical insights for microtransit deployment in Arlington, and the proposed framework is readily transferable to other cities with similar microtransit services.

## 1. Introduction

Microtransit, an emerging form of Demand Responsive Transit (DRT) service, bridges the gap between fixed-route transit and ride-hailing services by offering flexible, on-demand transportation in dynamically routed shuttles or vans (Slosky et al., 2022; Volinski et al., 2019; Yoon et al., 2022). In this ridepooling system, adaptable routes and schedules enable multiple passengers to share a vehicle when their travel plans align closely in a spatiotemporal sense (Daganzo & Ouyang, 2019). Empirical studies have highlighted the potential of microtransit in enhancing mobility for underserved populations (Erdoğan et al., 2024), replacing fixed-route public transit in rural residential areas (Hansen et al., 2021), and relieving traffic congestion by reducing the usage of private vehicles (Kawagughi et al., 2017). Despite these promising benefits, the financial viability of microtransit remains precarious. The reliance on public subsidies and short-term pilot funding requires intelligent revenue management strategies to achieve long-term profitability (Ghimire et al., 2024).

Recent studies on microtransit revenue management highlight the critical role of fare structures and pricing strategies (Erhardt et al., 2023; Geržinič et al., 2025; Pettine et al., 2022; Saha et al., 2025). Microtransit fares are suggested to be 1.5 to 2 times of fixed-route transit fares to leverage the premium value of on-demand convenience (Gonzales & Christofa, 2021), while a growing number of studies focus on targeted pricing policies such as means-tested discounts, subscription programs, fare capping, and context-specific subsidies (Hyland et al., 2025; Saha et al., 2025). However, empirical evidence reveals substantial heterogeneity in system performance; per-trip subsidies, cost recovery, and ridership responses vary significantly across different urban densities and service designs (Erhardt et al., 2023; He & Ma, 2022). These findings underscore the need for pricing strategies that account for behavioral heterogeneity, recognizing that travelers' preferences and willingness-to-pay (WTP) differ widely across geographic and sociodemographic contexts.

Discrete choice models (DCMs) and simulation-based approaches are widely used to forecast travel demand, capture traveler-operator interactions, and measure system performance (Caros & Chow, 2021; Rossetti et al., 2023; Yan et al., 2019). DCMs provide insights into travelers' sensitivities to key attributes such as travel time, fares, and transfers (Bowman & Ben-Akiva, 2001), while simulation frameworks integrate these behavioral models with vehicle routing and agent-based operations to assess microtransit performance under diverse policy scenarios (Costa et al., 2021; Namdarpour et al., 2024). However, three limitations in existing approaches continue to hinder effective microtransit revenue management. First, the reliance on survey data with limited sample size may lead to less representative patterns of microtransit usage (Cavallaro et al., 2023). Second, parametric assumptions on the distribution of taste parameters may fail to capture heterogeneous user preference (Hess, 2010; Sarrias, 2020). Third, existing studies largely overlook ride-pass subscription programs, which are critical pricing structures for microtransit operators (Marino & Jayakrishnan, 2024). Although advanced choice models and microtransit simulators have emerged with the increasing availability of large-scale travel data (Rath et al., 2023; Swait, 2022), their application to microtransit choice modeling remains limited and understudied.

To address the gap, we propose a novel approach that integrates citywide travel behavioral patterns into microtransit revenue management. The approach is distinguished by three key features: (1) leveraging citywide synthetic trip data to capture large-scale, fine-grained travel behavior, (2) estimating a nonparametric nested logit model to jointly analyze travel mode and ride-pass subscription decisions, and (3) employing a simulation-based framework to evaluate revenue outcomes and traveler welfare under diverse pricing scenarios.

We apply our framework to Via Arlington, a microtransit service operated in partnership with Via Transportation that serves as the sole public transit option in Arlington, TX. Launched in 2017 as a pilot with just 18 vehicles, the service has since grown into a citywide system with a fleet of approximately 70 vehicles covering over 100 square miles, making it one of the largest microtransit deployments in the United States (City of Arlington, 2024). The choice model is estimated using synthetic trip data from Replica Inc., and a microtransit simulator developed by Namdarpour et al. (2024) is calibrated to measure service performance given ride requests. We focus on ride-pass subscription programs and context-specific subsidies, and construct performance metrics including system revenue, microtransit ridership, and total ridepass subscribers for policy evaluation. These metrics jointly capture the operator, demand, and welfare dimensions of policy impact, offering actionable insights for service planning and subsidy design. Moreover, policy lessons drawn from Arlington's mature deployment can inform many emerging microtransit systems, such as RIDE in Wilson, NC[1], Metro Flex in Omaha, NE[2], and SAM On Demand in Sioux Falls, SD[3].

The remainder of the paper is organized as follows: Section 2 introduces policy instruments for microtransit revenue management and existing quantitative approaches, pointing out the current research gap and our contributions. Section 3 introduces our methodology of integrating choice modeling and microtransit simulation into a policy evaluation framework. Section 4 shows the results of an empirical study of Via microtransit in Arlington, TX. The final section concludes the key findings and discusses future work.

## 2. Literature review

### 2.1. Policy instruments for microtransit revenue management

A growing body of literature has examined the role of fare structures and pricing strategies in microtransit revenue management (Erhardt et al., 2023; Geržinič et al., 2025; Pettine et al., 2022; Saha et al., 2025). Microtransit fares are often benchmarked against existing fixed-route transit, typically ranging between 1.5 and 2 times the base fare (Gonzales & Christofa, 2021), in order to avoid diverting riders from traditional public transit while still capitalizing on the premium value of on-demand convenience. More recent studies have extended this discussion by simulating alternative fare structures, as well as evaluating their impacts on ridership, revenue, and vehicle miles traveled (Hyland et al., 2025; Saha et al., 2025). For example, Hyland et al. (2025) compared various fare levels of $2, $5, and $10 with dynamic pricing schemes (Sayarshad and Chow, 2015) to explore how pricing strategies can be optimized to improve financial sustainability while maintaining microtransit's role as a complement, rather than a substitute, to fixed-route transit. Dong et al. (2022) introduced a microtransit routing strategy that integrates different pricing structures, including fixed fares, distance-based fares, and zone-based fares, and found that zone-based fares are most effective in optimizing revenue and ridership. Other fare or tariff structures are summarized in Chin et al. (2016).

[1] https://www.wilsonnc.org/residents/all-departments/public-works/wilson-transit-ride-wilson-industrial-air-center/ride
[2] https://www.ometro.com/rider-guide/metro-flex/
[3] https://siouxareametro.info/on-demand

In addition to flat fare levels, existing studies have emphasized the importance of specialized fare policies such as means-tested discounts, subscription programs, fare capping, and context-specific subsidies (Darling et al., 2021; Pettine et al., 2022; Yuan et al., 2025). Means-tested discount programs are widely deployed to promote transit equity by offering reduced fares to low-income groups, seniors, or persons with disabilities (Darling et al., 2021). Subscription adoption (e.g., weekly or monthly ride-pass) can improve microtransit usage through price anchoring and habit formation (Marino & Jayakrishnan, 2024). Fare capping policies, whereby riders are provided unlimited rides once a daily or monthly spending threshold is reached, are increasingly recognized as effective strategies to contain rider expenditures while incentivizing continued use (Pettine et al., 2022). Context-specific subsidies, targeted toward special events or underserved geographic zones, have been shown to reduce congestion during peak periods and improve access to essential services such as healthcare (Yuan et al., 2025). Moreover, an emerging strand of studies focuses on the integration of microtransit fare policy into broader multimodal networks. For instance, Geržinič et al. (2025) argue that isolated fare adjustments have limited effects unless paired with complementary pricing instruments, particularly parking pricing and congestion charges.

Despite the flood of innovative ideas, empirical findings reveal substantial heterogeneity in outcomes across different contexts. For example, Erhardt et al. (2023) conducted multiple microtransit case studies and demonstrated that per-trip subsidy and cost recovery can vary drastically when operating in varying urban densities, highlighting that no single pricing or service strategy performs consistently well across all settings. Similarly, He and Ma (2022) and Ma et al. (2021) use data from the Kussbus pilot in the Luxembourg–Belgium cross-border region to show that traveler choices are highly sensitive to local service attributes such as walking distance, in-vehicle time, and fare levels. These findings indicate that microtransit pricing cannot rely on a "one-size-fits-all" model. Instead, designing effective pricing strategies requires a comprehensive understanding of behavioral heterogeneity, where individual preferences and willingness-to-pay (WTP) differ substantially across geographic and sociodemographic segments.

### 2.2. Choice models and simulation approaches for revenue analysis

Discrete choice models (DCMs) and simulation-based approaches have been widely used to forecast travel demand, capture complex rider-operator interactions, and evaluate the implications of pricing policies on system performance and revenue outcomes. In DCMs, travelers are assumed to make choices to maximize the overall utility they can expect to gain (Bowman & Ben-Akiva, 2001). Existing studies estimate DCMs to quantify travelers' tastes and sensitivities toward key service attributes, including travel time, trip fares, and transfer requirements. For example, Yan et al. (2019) analyzed ride-sourcing adoption at the University of Michigan, linking income and car ownership to mode preferences. Monchambert (2020) assessed the attractiveness of an innovative carpooling service by comparing it with buses, trains, and cars, evaluating key attributes for both drivers and passengers. Rossetti et al. (2023) designed an SP survey to understand travelers' preferences for first-mile/last-mile microtransit service in four cities in the U.S.: Miami, Minneapolis-Saint Paul, Washington, D.C., and Seattle. Choice probabilities and consumer surplus derived from DCMs can serve as key performance metrics in policymaking, enabling the evaluation of expected revenue outcomes and broader welfare impacts (Bills et al., 2022; Ren et al., 2024). Dong et al. (2022) incorporated the choice behavior within a dial-a-ride optimization problem to explore different pricing structures.

Simulation-based frameworks are effective in evaluating microtransit systems under varying operational policies (Jung & Chow, 2019; Markov et al., 2021; Yoon et al., 2022). These frameworks typically employ vehicle routing algorithms and agent-based simulations to assess microtransit performance using traveler-centric metrics (e.g., waiting time, in-vehicle time, rejection rates), operator-centric profits (e.g., operation cost and total revenue), or societal outcomes (e.g., mode shift and social welfare) (Calabrò et al., 2023; Leffler et al., 2024). For instance, Namdarpour et al. (2024) developed a non-myopic microtransit simulator that optimizes vehicle dispatching and passenger assignments to evaluate service performance. Although a majority of simulation approaches focus on "within-day" fleet dynamics by mimicking only the real-time vehicle dispatching and rider matching under fixed demand (Costa et al., 2021), there is a growing number of studies that consider dynamic adaptation or "day-to-day" adjustment to capture the equilibrium between demand and supply sides (Djavadian and Chow, 2017; Caros & Chow, 2021). These frameworks have emerged as critical tools by explicitly capturing the interplay between system state and traveler behavior.

However, several critical limitations in existing approaches continue to hinder effective microtransit revenue management. First, current studies mainly rely on SP surveys with limited sample sizes (Cavallaro et al., 2023). These surveys often overrepresent urban, tech-savvy early adopters while underrepresenting low-income, rural, or disabled travelers. For instance, a study of SmaRT Ride service in Sacramento revealed that low-income users faced such barriers as smartphone access and language differences, which were underrepresented in initial SP surveys (Xing et al., 2024). Second, choice models for microtransit often oversimplify user preferences by assuming homogeneous or parametric taste distribution. In the mixed logit (MXL) model, taste parameters are typically assumed to follow normal, uniform, or triangular distribution (McFadden & Train, 2000), which may not be a good fit given unobserved habitual factors (Hess, 2010; Sarrias, 2020). Although advanced nonparametric approaches—such as nonparametric mixed logit (Vij & Krueger, 2017), individual parameter logit (Swait, 2022), and agent-based mixed logit (Ren et al., 2025; Ren & Chow, 2022)—have been proposed, their application to microtransit choice modeling remains understudied. Third, most studies focus on travel mode and route choice, while overlooking subscription programs that significantly drive long-term revenue (Hörcher & Graham, 2020). Without linking lower-level travel behaviors to upper-level subscription decisions, policymakers risk missing opportunities to design strategies that discount individual rides in exchange for higher ridership, stronger willingness to subscribe, and long-term user retention.

### 2.3. Our contributions

Building on these limitations, we propose an integrated modeling framework that: (1) uses citywide synthetic population data to capture fine-grained spatiotemporal travel patterns, (2) estimates a nonparametric nested logit model to jointly analyze travel mode and ride-pass subscription choices, and (3) integrates a simulation-based evaluation to quantify revenue outcomes and traveler welfare under diverse pricing scenarios.

Methodologically, our approach differs from existing methods in three key aspects. First, we integrate synthetic trip data (1.35 million trips made by 0.4 million Arlington residents) with marginal subscription data to provide a citywide representation of microtransit demand, which is not feasible when relying solely on small-scale survey data. Second, the choice model captures taste heterogeneity using a nonparametric estimation approach (Ren & Chow, 2022), enabling behavioral analysis at the population, individual, and trip levels without relying on predefined distributional forms (van Cranenburgh et al., 2022). Third, a nested structure jointly models travel mode choice and ride-pass subscription decisions, with parameters from the lower-branch mode choice model transformed into individual-specific attributes in the upper-branch subscription model, enabling subscription behavior to be inferred even with limited pass data.

From a practical standpoint, the integrated framework is designed to support policy evaluation across several dimensions that are directly relevant to microtransit operators and transit agencies. The nested structure enables operators to quantify how changes in service performance (e.g., in-vehicle time, waiting time) propagate through mode choice into subscription uptake, providing a tool for assessing the revenue implications of both pricing and operational decisions jointly rather than in isolation. The simulation-based workflow further closes this loop by allowing supply-side attributes and demand-side behavioral responses to interact dynamically, capturing feedback effects (e.g., demand-induced service degradation) that static models cannot represent. This makes it possible to evaluate not only flat-fare and subscription pricing strategies but also context-specific interventions such as event-based and place-based subsidies. Moreover, because the framework is built on synthetic population data and a calibrated microtransit simulator, it is readily transferable to other cities with similar deployments, given locally calibrated inputs. We demonstrate the framework's applicability through a case study in Arlington, TX, and draw broader policy implications in Section 5.

## 3. Proposed methodology

The proposed methodology integrates joint subscription and mode choice modeling with dynamic microtransit simulations (Fig. 1). The lower-branch mode choice model employs an agent-based mixed logit (AMXL) approach to estimate travel mode decisions using citywide synthetic trip data. The upper-branch subscription choice model incorporates travelers' price sensitivities and microtransit utilities from the lower branch, calibrating general taste parameters to match the observed subscription levels from marginal subscription data. Model predictions are then embedded in a simulation-based framework that iteratively updates perceived service performance until reaching user equilibrium, enabling the evaluation of microtransit pricing strategies and their impacts on system revenue, mode shifts, and traveler welfare.

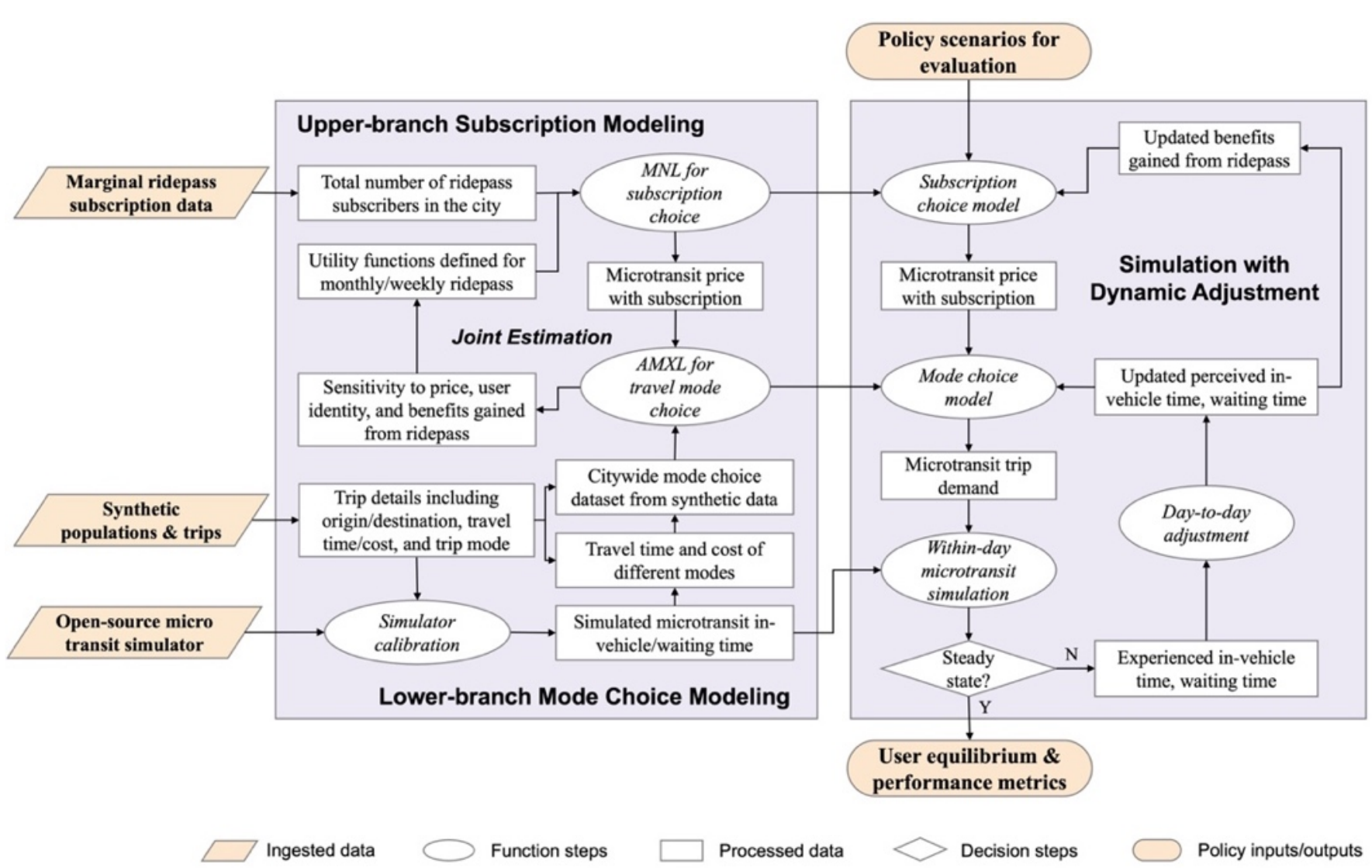

**Fig. 1.** The proposed framework.

### 3.1. Nested subscription and mode choice model

We propose a nested structure to estimate joint ride-pass subscription and travel mode choice. In the upper branch, individuals determine whether to purchase a weekly ride pass, a monthly ride pass, or no ride pass. By subscribing to a ride pass, travelers pay a fixed amount in advance and can take unlimited microtransit trips until the pass expires. Accordingly, subscription decisions are impacted by the cost of the ride pass and the utility gained from free microtransit trips after subscribing. The latter part comes from the lower branch, where individuals decide which mode to use conditional on the upper-branch decisions. In the lower branch, trips on weekdays and weekends are separated into two datasets. Two mode choice models are estimated to account for their divergent patterns. We assume individuals have different tastes on weekdays and weekends, and that weekday and weekend choices are independent of each other. The mode choice set includes traditional modes (e.g., driving, walking, biking, etc.) as well as microtransit. Factors such as travel time, trip purpose, tour type, and cost conditional on ride pass ownership impact the mode choice. The nested structure of our model is shown in Fig. 2. Notations used in the model are listed in Appendix Table A1.

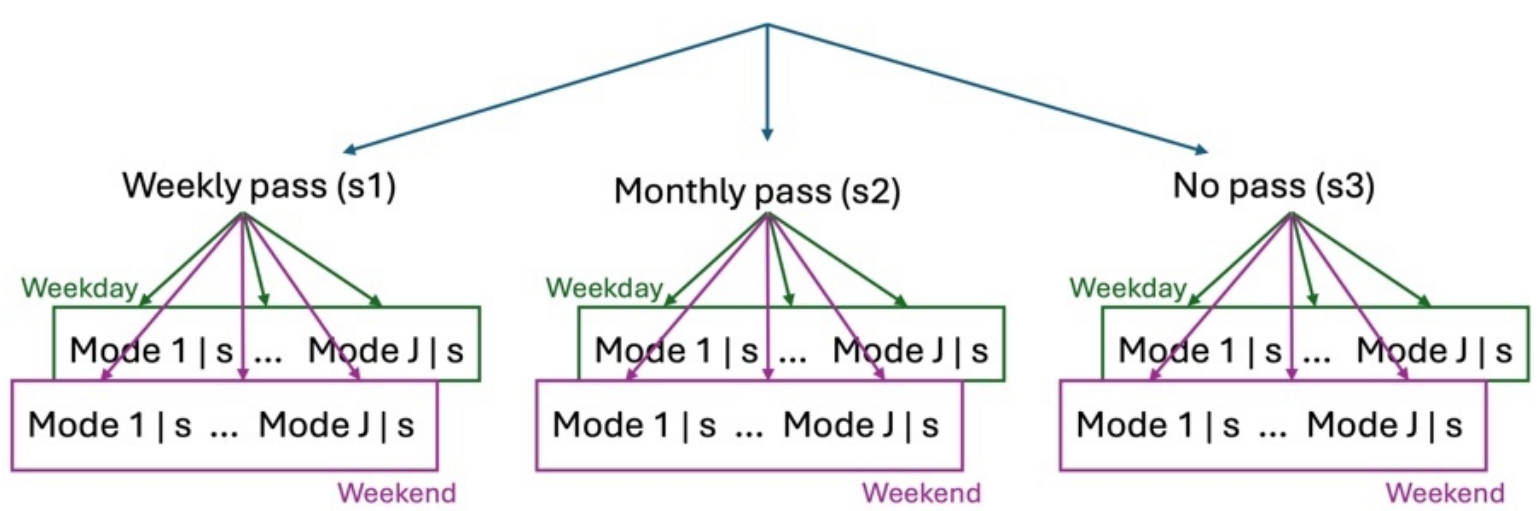

**Fig. 2.** The nested structure in the model.

#### *3.1.1. Upper-branch subscription choice model*

We consider three alternatives according to current subscription programs: subscribing to a weekly ride pass, subscribing to a monthly ride pass, and opting not to subscribe to a ride pass. From a weekly perspective, the utility of individual $n$ choosing these alternatives are defined in Eqs. (1) – (3).

$$U_{n,weekly} = \beta_{price,n} Price_{weekly} + \beta_{gain}^{we} Gain_n^{we} + \beta_{gain}^{wd} Gain_n^{wd} + \beta_{MT-user} MT_n + asc_{weekly} + \varepsilon_{n,weekly}, \quad \forall n \in N \tag{1}$$

$$U_{n,monthly} = \beta_{price,n} Price_{monthly}/4 + \beta_{gain}^{we} Gain_n^{we} + \beta_{gain}^{wd} Gain_n^{wd} + \beta_{MT-user} MT_n + asc_{monthly} + \varepsilon_{n,monthly}, \quad \forall n \in N \tag{2}$$

$$U_{n,none} = \varepsilon_{n,none}, \quad \forall n \in N \tag{3}$$

where $U_{n,weekly}$, $U_{n,monthly}$, and $U_{n,none}$ denote individual $n$'s utility per week by subscribing to a weekly, monthly, and no ride pass. $Price_{weekly}$ and $Price_{monthly}$ are the prices of the ride pass (the price of monthly ride pass is divided by 4 to obtain the cost per week). $\beta_{price,n}$ is a parameter reflecting individual $n$'s sensitivity to price. $Gain_n^{we}$ and $Gain_n^{wd}$ denote the benefits gained from free microtransit trips by holding a ride pass, which depend on individual $n$'s microtransit trip frequency on weekday ($we$) and weekend ($wd$). $\beta_{gain}^{we}$, $\beta_{gain}^{wd}$ are parameters reflecting the general sensitivity to these gained benefits. $MT_n$ is a binary variable indicating whether individual $n$ has experience riding microtransit before, and $\beta_{MT-user} > 0$ is its parameter assuming that individuals who have used microtransit have a higher probability to purchase ride pass. $asc_{weekly}$ and $asc_{monthly}$ are alternative specific constants of weekly and monthly ride pass. $\varepsilon_{n,weekly}$, $\varepsilon_{n,monthly}$, and $\varepsilon_{n,none}$ are independent and identically distributed (i.i.d.) random utilities following a Gumbel distribution. The probability of individual $n$ choosing subscription option $s$ can be computed using the multinomial logit (MNL) form.

The primary challenge in estimating this model as a stand-alone MNL model is that only marginal subscription data are available due to privacy concerns. This limitation makes attributes such as $Gain_n^{we}$, $Gain_n^{wd}$, $MT_n$ difficult to measure and renders the individual parameter $\beta_{price,n}$ infeasible to estimate. However, it is possible to obtain this information from a lower-branch mode choice model estimated using large-scale travel data from synthetic populations/trips and advanced estimation approaches. In that case, $Gain_n^{we}$, $Gain_n^{wd}$, $MT_n$, and $\beta_{price,n}$ can be obtained from the choice observations and taste parameters from the lower branch, as shown in Eqs. (4) – (6).

$$Gain_n^d = \sum_{t \in T_n^d} \left( \ln \sum_{j \in J} e^{V_{n,j,t}^{rp}/\beta_{gain}^d} - \ln \sum_{j \in J} e^{V_{n,j,t}^{nrp}/\beta_{gain}^d} \right), \qquad \forall n \in N, d \in \{we, wd\} \tag{4}$$

$$MT_n = 1\left(\sum_{t \in T} 1(j_{n,t} = MT) > 0\right), \ \forall n \in N \tag{5}$$

$$\beta_{price,n} = f_{rp} \beta_{cost,n}, \ \forall n \in N \tag{6}$$

where $d \in \{we, wd\}$ indicates the day of the week, $T_n^d$ is the set of trips made by individual $n$ on day $d$, $J$ is the mode choice in the lower branch, $V_{n,j,t}^{rp}$ represents the systematic utility with ride pass trip discounts, and $V_{n,j,t}^{nrp}$ represents the systematic utility without ride pass trip discounts. Eq. (4) indicates that the "benefits" gained from holding a ride pass are measured as the logsum utility or inclusive value (Ben-Akiva & Lerman, 1985) in the lower branch. The terms $\ln \sum_{j \in J} e^{V_{n,j,t}^{rp}/\beta_{gain}^d}$ and $\ln \sum_{j \in J} e^{V_{n,j,t}^{nrp}/\beta_{gain}^d}$ represent the inclusive values for the nests with and without ride pass, respectively. The parameters $\beta_{gain}^{we}$ and $\beta_{gain}^{wd}$ serve as two nesting parameters that scale the mode choice utilities on weekdays and weekends. Eq. (5) ensures that individual $n$ is classified as a microtransit user if the individual has at least one observed microtransit trip in the lower-branch choice dataset, where $1(.)$ is an indicator function that returns 1 if the condition in the parentheses holds and 0 otherwise. Eq. (6) links the individual $n$'s sensitivity to trip cost with the sensitivity to the ride pass price by introducing a generic parameter $f_{rp}$, which reflects the proportional relationship between individuals' valuation of per-trip costs and their willingness to pay for a ride pass.

### *3.1.2. Lower-branch mode choice model*

In the mode choice model, we differentiate the utility function of choosing microtransit from other modes (e.g., driving, public transit, biking, walking), conditional on the subscription plan choice. For each trip $t \in T$ made by individual $n \in N$ on a weekday or a weekend, the utility functions choosing microtransit and other modes are defined in Eqs. (7) – (8).

$$V_{n,MT,t|s} = \beta_{tt,n} TT_{MT,t} + \beta_{wt,n} WT_{MT,t} + Y_{n,none}\beta_{cost,n} COST_{n,MT,t} + \sum_{p\in P} \beta_p Purp_{n,p,t} + \sum_{q\in Q} \beta_q Tour_{n,q,t} + asc_{MT,t}, \quad \forall n \in N, t \in T \tag{7}$$

$$V_{n,j,t|s} = V_{n,j,t} = \beta_{tt,n} TT_{j,t} + \beta_{cost,n} COST_{j,t} + asc_{j,t}, \quad \forall n \in N, j \in J, j \neq MT, t \in T \tag{8}$$

where $V_{n,MT,t|s}$ is the systematic utility of choosing microtransit, which is conditional on individual $n$'s subscription choice $s$. $TT_{MT,t}$ denotes microtransit travel time, $WT_{MT,t}$ denotes microtransit waiting time, and $COST_{MT,t}$ denotes microtransit trip fare based on fare schedules. $Y_{n,none}$ is a binary variable indicating if individual $n$ does not subscribe to a ride pass at the upper branch. Eq. (7) ensures that when $Y_{n,none} = 0$, the microtransit utility corresponds to a ride pass scenario, $V_{n,j,t}^{rp}$, and when $Y_{n,none} = 1$, the microtransit utility corresponds to the non-ride pass scenario, $V_{n,j,t}^{nrp}$. Eq. (8) defines the systematic utility of choosing other mode $j \in J$. We include several dummy variables to consider the effects of trip purpose and tour type on the preference of microtransit over other modes. $Purp_{n,p,t}$ is a dummy variable indicating whether trip $t$ made by individual $n$ serves purpose $p$. $Tour_{n,q,t}$ is a dummy variable indicating whether trip $t$ made by individual $n$ is part of a tour with type $q$. $P$ is the set of trip purposes such as working or shopping. $Q$ is the set of such tour types as commuting tour or home-based tour.

$\beta_{tt,n}$, $\beta_{wt,n}$, and $\beta_{cost,n}$ represent travel time and cost parameters, which are assumed to vary across individuals. $asc_{MT,t}$ and $asc_{j,t}$ denote alternative-specific constants, which are assumed to vary across trip origin-destination (OD) pairs. $\beta_p, p \in P$ and $\beta_q, q \in Q$ are parameters of trip purpose and tour type, which are assumed to be homogeneous across individuals and trips. By allowing a nonparametric distribution at different levels, we account for the influence of sociodemographic and built-environment factors without requiring handcrafted model specifications where modelers have to determine how to specify each interaction term prior to model estimation (van Cranenburgh et al., 2022). The model makes sense when (1) the choice dataset is sufficiently large, and (2) we only want to make predictions for the same population. These conditions are satisfied when the model is estimated using citywide synthetic trip data.

Based on the nested structure, the probability of individual $n$ choosing subscription option $s$ and using mode $j$ in trip $t$ can be defined in Eqs. (9) – (11).

$$P_{n,s,j,t} = P_{n,s} P_{n,j,t|s}, \quad \forall n \in N, s \in S, j \in J, t \in T \tag{9}$$

$$P_{n,s} = \frac{e^{V_{n,s}}}{\sum_{s\in S} e^{V_{n,s'}}}, \forall n \in N, s \in S \tag{10}$$

$$P_{n,j,t|s} = \frac{e^{V_{n,j,t|s}}}{\sum_{j\in J} e^{V_{n,j',t|s}}}, \quad \forall n \in N, s \in S, j \in J, t \in T \tag{11}$$

where $S$ is the choice set of ride pass options and $J$ is the choice set of travel modes. $P_{n,s}$ denotes the probability in the upper branch and $P_{n,j,t|s}$ denotes the probability in the lower branch. The estimation algorithm for the nested subscription–mode choice model is introduced in Appendix A2.

## 3.2. Microtransit simulation framework

### *3.2.1. Within-day microtransit simulator*

A key part of the framework is the within-day microtransit simulator, which is designed to simulate trip-level microtransit in-vehicle and waiting times based on incoming trip requests, while considering congestion and queueing effects. We use an open-source simulator developed by Namdarpour et al. (2024), NOMAD-RPS, which takes microtransit trip requests as input and outputs service performance metrics by determining which vehicles to dispatch and transport passengers from their origins to their destinations. We selected NOMAD-RPS for two reasons: (1) it is open source with a well-maintained GitHub repository (https://github.com/BUILTNYU/ridepooling), and (2) it employs a parametric non-myopic cost function approximation policy to sequentially find the best matches between riders and vehicles, which can be calibrated to fit real-world fleet operations.

Several parameters should be calibrated to ensure the simulator accurately reflects a specific microtransit system. These parameters include number of vehicles, maximum passenger walking time, maximum passenger waiting time, the balance between operator and user cost, and the degree of lookahead approximation employed. Calibration of these parameters can be conducted by comparing simulation outputs with observed performance metrics from real-world microtransit services, such as average in-vehicle time, average waiting time, and average vehicle occupancy.

### *3.2.2. Dynamic adjustment process*

A day-to-day process characterizes the dynamics in adjustments made by both travelers and operators each day as a dynamic system (Rath et al., 2023). Djavadian and Chow (2017) showed that such adjustment processes can reach a stochastic user equilibrium (SUE) to simulate on-demand systems. However, existing studies have neither incorporated nonparametric mode choice parameters into microtransit demand estimation nor considered ride pass subscription choices at a higher level.

The process introduced by Rath et al. (2023) is modified to address the requirements of our framework. In our process, only travelers are allowed to adjust their behavior, whereas prices and operational decisions from the operator side are treated as explicit attributes. The adjustment process is also modified to incorporate both mode and subscription choices, such that perceived microtransit in-vehicle and waiting times not only influence mode selection, but also determine the benefits derived from ride pass and consequently affect ride pass ownership. The right part of Fig. 1 illustrates the whole process.

At the start of the simulation, a policy scenario is defined as the input, specifying operational decisions such as ride pass pricing, fare schedules, and fleet size. The microtransit fare for each trip is then calculated based on trip origins and destinations as well as predicted ride pass ownership. Travelers subsequently make their mode choices by evaluating trip fares and perceived service attributes, including in-vehicle and waiting times. These mode decisions generate the microtransit trip requests in the within-day microtransit simulator. Microtransit in-vehicle and waiting times are updated from day to day. A traveler $n$ who used microtransit on day $d$ learns from his/her experience on the day and updates their perceived in-vehicle and waiting times with a learning rate $\theta$, as shown in Eqs. (12) – (13).

$$TT_{MT,n}^{d+1} = (1-\theta)TT_{MT,n}^{d} + \theta ETT_{MT,n}^{d}, \qquad \forall n \in N_{MT} \tag{12}$$

$$WT_{MT,n}^{d+1} = (1-\theta)WT_{MT,n}^{d} + \theta EWT_{MT,n}^{d}, \qquad \forall n \in N_{MT} \tag{13}$$

where $N_{MT}$ is the set of microtransit users. $TT_{MT,n}^{d}$ and $WT_{MT,n}^{d}$ denote perceived microtransit in-vehicle time and waiting time for user $n$ at the beginning of day $d$. $ETT_{MT,n}^{d}$ and $EWT_{MT,n}^{d}$ represent experienced microtransit in-vehicle time and waiting time obtained from the microtransit simulator. For travelers who did not use microtransit, their perceived times are updated with the population's average perceived times which is the successive average of average times in the past days. For instance, the population's perceived in-vehicle time on day $d$ ($\overline{TT}_{MT}^{d}$) is defined in Eqs. (14) – (15).

$$\overline{TT}_{MT}^{d} = \left(1-\frac{1}{d}\right)\overline{TT}_{MT}^{d-1} + \frac{1}{d}E\overline{TT}_{MT}^{d} \tag{14}$$

$$E\overline{TT}_{MT}^{d} = \frac{1}{|N_{MT}|}\sum_{n\in N_{MT}} ETT_{MT,n}^{d} \tag{15}$$

At the end of each day (or iteration), we update $Gain_{n}^{we}$, $Gain_{n}^{wd}$, $MT_{n}$, and the ride pass ownership prediction, and then check whether the system has reached a steady state. The adjustment stops when the change of microtransit ridership and ride pass ownership is below 1%. We recall the proposition from Djavadian and Chow (2017) where the agent-based day-to-day process with a Method of Successive Averages (MSA) converges almost surely to a stochastic user equilibrium (SUE).

### 3.3. Performance metrics for policy evaluation

#### *3.3.1. Consumer surplus and elasticity*

Based on the parameters from the behavioral model, we calculate the *consumer surplus* (Chipman & Moore, 1980) of subscription choice and mode choice, as shown in Eqs.(16) – (17).

$$CS_{n,S} = -\frac{1}{\beta_{price,n}}\ln\left(\sum_{s\in S} e^{V_{n,s}}\right) + C_{n,S}, \qquad \forall n \in N \tag{16}$$

$$CS_{n,J,t}^{d} = -\frac{1}{\beta_{cost,n}^{d}}\ln\left(\sum_{j\in J} e^{V_{n,j,t}^{d}/\beta_{gain}^{d}}\right) + C_{n,J}^{d}, \qquad \forall n \in N, d \in \{we, wd\} \tag{17}$$

where $CS_{n,S}$ represents the consumer surplus of subscription choice. One over the ride pass price parameter $-\frac{1}{\beta_{price,n}}$ is used to convert it into monetary units (dollars). $CS_{n,J,t}^{d}$ represents the consumer surplus of mode choice on weekdays or weekends. One over the travel cost parameter $-\frac{1}{\beta_{cost,n}^{d}}$ is used to convert it into monetary units (dollars). $C_{n,S}$ and $C_{n,J}^{d}$ are arbitrary constants, so the absolute surplus is less meaningful than surplus differences between scenarios. The value of ride pass plans and microtransit service can be measured using the difference in consumer surplus between two scenarios: one with these alternatives and one without, as shown in Eqs. (18) – (19).

$$CV_{n,S\rightarrow S^-} = CS_{n,S} - CS_{n,S^-}, \qquad \forall n \in N \tag{18}$$

$$CV^d_{n,J\rightarrow J^-,t} = CS^d_{n,J,t} - CS^d_{n,J^-,t}, \qquad \forall n \in N, d \in \{we, wd\} \tag{19}$$

where $CV$ denotes the compensating variation—the amount an individual should be compensated to be as well off as before a scenario change (Small & Rosen, 1981). $S$ is the subscription choice set that includes ride pass options, while $S^-$ is the set that only includes the no-ride-pass option. $J$ is the mode choice set that includes microtransit, while $J^-$ is the set excluding microtransit.

Moreover, we calculate the *elasticity* of ride pass ownership (weekly pass and monthly pass) with respect to microtransit service performance (in-vehicle time and waiting time on weekdays and weekends). For instance, the elasticity of an individual $n$'s probability of purchasing weekly ride pass ($P_{n,weekly}$) w.r.t. microtransit waiting time of trip $t$ on weekday ($WT^{we}_{MT,t}$) is defined in Eqs.(20) – (22).

$$E^{P_{n,weekly}}_{WT^{we}_{MT,t}} = \frac{\partial P_{n,weekly}}{\partial Gain^{we}_n} \cdot \frac{\partial Gain^{we}_n}{\partial WT^{we}_{MT,t}} \cdot \frac{WT^{we}_{MT,t}}{P_{n,weekly}}, \qquad \forall n \in N, t \in T \tag{20}$$

$$\frac{\partial P_{n,weekly}}{\partial Gain^{we}_n} = P_{n,weekly}\left(1 - P_{n,weekly}\right)\beta^{we}_{gain}, \qquad \forall n \in N \tag{21}$$

$$\begin{aligned} \frac{\partial Gain^{we}_n}{\partial WT^{we}_{MT,t}} &= \frac{\partial \sum_{t\in T^{we}_n}\left(\ln\left(\sum_{j\in J} e^{V^{rp}_{nj,t}}\right) - \ln\left(\sum_{j\in J} e^{V^{rp}_{nj,t}}\right)\right)}{\partial WT^{we}_{MT,t}} \\ &= \frac{1}{\sum_{j\in J} e^{V^{rp}_{nj,t}}} \cdot e^{V^{rp}_{n,MT,t}} \cdot \beta^{we}_{wt,n} - \frac{1}{\sum_{j\in J} e^{V^{nrp}_{nj,t}}} \cdot e^{V^{nrp}_{n,MT,t}} \cdot \beta^{we}_{wt,n} \\ &= \beta^{we}_{wt,n}\left(P^{rp}_{n,MT,t} - P^{nrp}_{n,MT,t}\right), \qquad \forall n \in N, t \in T \end{aligned} \tag{22}$$

where $E^{P_{n,weekly}}_{WT^{we}_{MT,t}}$ represents the elasticity, $Gain^{we}_n$ represents the "benefits" gained from holding a ride pass on weekdays, $\frac{\partial P_{n,weekly}}{\partial Gain^{we}_n}$ represents the partial derivative of $P_{n,weekly}$ w.r.t. $Gain^{we}_n$ derived from the upper-branch model, and $\frac{\partial Gain^{we}_n}{\partial WT^{we}_{MT,t}}$ represents the partial derivative of $Gain^{we}_n$ w.r.t. $WT^{we}_{MT,t}$ derived from the lower-branch model. $P^{rp}_{n,MT,t}$ and $P^{nrp}_{n,MT,t}$ represent the probability of choosing microtransit with and without a ride pass. $\beta^{we}_{wt,n}$ is individual $n$'s parameter of microtransit waiting time on weekdays.

*3.3.2. Microtransit ridership, number of subscribers, and system revenue*

Once the simulation reaches a steady state, several key metrics are calculated for policy evaluation. *Daily microtransit ridership (DMR)* is computed by summing the trip-level probabilities of choosing microtransit. To derive the daily microtransit ridership (trips/day), we average weekday and weekend ridership based on the respective number of days in a week, as shown in Eq. (23).

$$DMR = \frac{5}{7}\sum_{n\in N}\sum_{t\in T^{we}} Y_{n,MT,t} + \frac{2}{7}\sum_{n\in N}\sum_{t\in T^{wd}} Y_{n,MT,t} \tag{23}$$

where $Y_{n,MT,t}$ is a binary variable indicating whether individual $n$ chooses microtransit in trip $t$, $T^{we}$ is the set of weekday trips, and $T^{wd}$ is the set of weekend trips.

*Total number of subscribers (TNS)* is computed by summing the number of individuals subscribing weekly and monthly ride pass, as shown in Eq. (24). The unit is persons/week.

$$TNS = NS_{weekly} + NS_{monthly}/4 = \sum_{n\in N} Y_{n,weekly} + \sum_{n\in N} Y_{n,monthly}/4 \tag{24}$$

where $Y_{n,weekly}$ and $Y_{n,monthly}$ are binary variables indicating whether individual $n$ purchases a weekly ride pass or a monthly ride pass, respectively.

*Total daily revenue (TDR)* is measured in units of dollars per day ($/day) and consists of revenue collected from trip fares and ride pass subscriptions, as shown in Eq. (25). To calculate trip fare revenue, we only consider microtransit trips made by non-subscribers since microtransit service is free to ride pass subscribers. Daily ride pass revenue is calculated by dividing weekly ride pass revenue by 7 and monthly ride pass revenue by 30, and then summing these two amounts.

$$\begin{aligned} TDR = {} & \frac{5}{7}\sum_{n\in N_{none}}\sum_{t\in T^{we}} COST_{MT,t}Y_{n,MT,t} + \frac{2}{7}\sum_{n\in N_{none}}\sum_{t\in T^{wd}} COST_{MT,t}Y_{n,MT,t} \\ & + \frac{1}{7}Price_{weekly}NS_{weekly} + \frac{1}{30}Price_{monthly}NS_{monthly} \end{aligned} \tag{25}$$

where $N_{none}$ is the set of non-subscribers, $COST_{MT,t}$ denotes the microtransit trip fare based on fee schedules, $Price_{weekly}$ denotes the price of weekly ride pass, and $Price_{monthly}$ denotes the price of monthly ride pass.

## 4. Case study: Via Microtransit in Arlington, TX

In this section, we apply the proposed methodology to a case study in Arlington, TX. With a population of 0.4 million, Arlington is one of the largest U.S. cities without a fixed-route public transit system. In December 2017, the city launched a public-private partnership for microtransit operated by Via, making it Arlington's sole public transit option, and the first city in the U.S. to adopt microtransit as the primary mode of public transport (Via, 2020). The service has expanded annually, and in 2021, the city contracted Via to operate approximately 70 six-passenger minivans covering the entire city. Travelers can summon a vehicle with a 12-15 minute waiting time and pay a fare ranging from $3 to $5 per trip (based on trip distance) or subscribe to a weekly ride pass ($25 per week) or a monthly ride pass ($80 per month). These features make Via Arlington an ideal case study.

## 4.1. Setup of experiments

### *4.1.1. Data collection*

The synthetic population dataset provided by Replica Inc. comprises 1.35 million trips (0.70 million on a typical weekday and 0.65 million on a typical weekend) made by 0.4 million Arlington residents in Q2 2023. The dataset includes trip details such as origin and destination, mode, distance, duration, and cost. The dataset was generated using a combination of mobile location data, resident demographics, built environment data, and economic activity data (Replica Inc., 2024). According to Replica's nationwide data validation report (Replica Inc., 2024), the largest error in demographic attributes for a single census tract unit is within 5% compared to American Community Survey (ACS) data, while the largest error in commute mode share for a single census tract unit is within 10% compared to Census Transportation Planning Products (CTPP) data.

We define population groups using a hierarchical classification scheme. First, individuals aged 65 and older are identified as senior segment. Second, individuals enrolled in school are classified as student segment. Finally, the remaining population is divided into low-income segment and not–low-income segment according to the Federal Poverty Guidelines (ASPE, 2023). A comparison between Replica's data with the American Community Survey (ACS) data is provided in Appendix A3 Table A2.

Five trip modes are included: driving, biking, walking, carpool (trips made by several passengers in an auto vehicle), and microtransit. No trips are classified as public transit, as Arlington does not have a fixed-route public transit system. On the weekday, the mode shares of synthetic trips are as follows: driving accounts for 63.55%, biking for 0.85%, walking for 8.64%, carpooling for 26.48%, and microtransit for 0.48%. On the weekend, the mode shares of synthetic trips are as follows: driving accounts for 67.50%, biking for 0.43%, walking for 6.11%, carpooling for 25.59%, and microtransit for 0.36%. A visualization of census block group-level trip OD pairs can be found in Appendix A3 Table A1.

In addition, microtransit service data for May 2023 are provided by City of Arlington (CoA), which align well with Replica's synthetic data for the corresponding time period. Due to data privacy concerns, only aggregated values are available, including the total number of monthly and weekly subscribers, as well as the average microtransit in-vehicle time, waiting time, and utilization rate (measured as the number of served passengers per vehicle per hour). We use the average service performance to calibrate our microtransit simulator and the total number of subscribers to estimate the upper-branch model.

### *4.1.2. Microtransit simulator calibration*

We employ a grid search with predefined searching ranges and intervals to calibrate parameters in NOMAD-RPS, ensuring that the simulation results closely align with the average service performance provided by CoA. The calibration process focuses on several parameters mentioned in Section 3.2.1.

- The number of vehicles in operation is calibrated within a range of 40 to 80, with an interval of 5, and the final values set at 55 for weekdays and 35 for weekends.
- The maximum passenger walking time is explored within a range of 10 to 20 minutes, with an interval of 2 minutes, and the final value set at 12 minutes for both weekdays and weekends.

- The maximum passenger waiting time is tuned between 30 and 60 minutes, with an interval of 10 minutes, and the final value set at 40 minutes for both weekdays and weekends.
- The weight assigned to operating cost (Theta) is calibrated between 0 and 1, with an interval of 0.1, and the final value set at 0.2 for both weekdays and weekends.
- The forward-looking parameter (Beta) is calibrated between 0 and 1, with an interval of 0.1, and the final values set at 0.2 for weekdays and 0.1 for weekends.

To accelerate the simulation for the entire city, we aggregate trip origins and destinations to their respective block group centroids and use the links between these centroids to construct the network, resulting in a computing time of approximately one minute per simulation round. Table 1 presents a comparison between simulation results and the data from CoA. The simulated average in-vehicle travel time is slightly longer than the observed values, while the simulated utilization rate is slightly lower. The largest percentage difference is approximately 15%. A sensitivity analysis examining how this level of error propagates through the downstream mode choice and ride-pass predictions is provided in Section 4.2.4.

**Table 1**
A comparison between simulation results and ground truth data.

| | Simulation results | Data from CoA | % Difference |
|---|---|---|---|
| Average in-vehicle time (weekday) | 19.99 min | 17.31 min | +15.48% |
| Average in-vehicle time (weekend) | 18.76 min | 16.44 min | +14.12% |
| Average utilization rate (weekday) | 3.467 | 3.637 | –4.67% |
| Average utilization rate (weekend) | 3.682 | 4.017 | –8.34% |
| Average waiting time (weekday) | 14.11 min | 12-15 min | -- |
| Average waiting time (weekend) | 11.71 min | 12-15 min | -- |

Note: utilization rate is calculated as average number of served passengers per vehicle per hour.

### *4.1.3. Model specification*

In the upper branch, the prices of weekly and monthly ride passes are set at $25 per week and $80 per month, in accordance with Via Arlington's current subscription pricing policy. Marginal subscription data from CoA shows that in May 2023, there were 412 active subscribers, accounting for 0.14% of the total residents (0.29 million) who had at least one synthetic trip. Among these subscribers, 320 opted for the weekly ride pass, while 92 subscribed to the monthly ride pass. Since we are building the choice model using the total population from the synthetic data, we can directly set $Num_{weekly} = 320$ and $Num_{monthly} = 92$ to estimate subscription choice parameters.

In the lower branch, since Replica's synthetic trip data combines microtransit in-vehicle and waiting time, we obtain $WT_{MT,t}$ and $TT_{MT,t}$ from the calibrated simulator based on observed microtransit demand. The utility functions of the five modes are defined in Eqs. (26) – (30).

$$V_{n,driving,t} = \beta_{auto_tt,n} TT_{driving,t} + \beta_{cost,n} COST_{driving,t} + asc_{driving,t}, \quad \forall n \in N, t \in T \tag{26}$$

$$V_{n,biking,t} = \beta_{non_auto_tt,n} TT_{biking,t} + asc_{biking,t}, \quad \forall n \in N, t \in T \tag{27}$$

$$V_{n,walking,t} = \beta_{non_auto_tt,n} TT_{walking,t} + asc_{walking,t}, \quad \forall n \in N, t \in T \tag{28}$$

$$V_{n,carpool,t} = \beta_{auto_tt,n} TT_{carpool,t} + \beta_{cost,n} COST_{driving,t}, \qquad \forall n \in N, t \in T \tag{29}$$

$$\begin{aligned} V_{n,MT,t|s} = \beta_{auto_tt,n} TT_{MT,t} + \beta_{wt,n} WT_{MT,t} + Y_{n,none} \beta_{cost,n} COST_{n,MT,t} \\ + \beta_{shopping} SH_{n,t} + \beta_{school} SC_{n,t} + \beta_{other} OT_{n,t} \\ + \beta_{commute} COM_{n,t} + \beta_{noncom} NCOM_{n,t} + asc_{MT,t}, \\ \forall n \in N, t \in T \end{aligned} \tag{30}$$

where $\beta_{auto_tt,n}$, $\beta_{non_auto_tt,n}$ are parameters of auto and non-auto travel time. $TT_*$, $COST_*$, and $asc_*$ present the travel time, travel cost, and mode-specific constants for these modes. Carpool is set as the reference level. $SH_{n,t}$ is a dummy variable indicating whether the trip is for shopping, $SC_{n,t}$ is a dummy variable indicating whether the trip is to school, and $OT_{n,t}$ is a dummy variable indicating whether the trip serves other purposes. $COM_{n,t}$ is a dummy variable indicating whether the trip is part of a commute tour, and $NCOM_{n,t}$ is a dummy variable indicating whether the trip is part of a non-commute tour.

One of the enduring challenges in building a RP choice dataset is that only the attributes of the chosen mode are available, i.e., if an individual has an observed driving trip, we need to infer the travel time and cost of microtransit or biking for that trip. Using large-scale synthetic trip data can address this challenge, as for each origin-destination (OD) pair, we have a sufficient number of trips using different modes. In that case, missing mode attributes in a trip can be obtained by averaging mode attributes from other observations along the same OD pair. Since microtransit trips account for only a small proportion, we infer $WT_{MT,t}$ and $TT_{MT,t}$ for non-microtransit users by averaging the corresponding values from microtransit users from the same origin to the same destination. The attributes are summarized in Table 2.

**Table 2**
Summary of mode attributes in the lower branch.

| Attribute | Notation | Weekday | | | Weekend | | |
|---|---|---|---|---|---|---|---|
| | | Mean | Std. | 50% | Mean | Std. | 50% |
| Driving travel time (min) | $TT_{driving,t}$ | 12.52 | 6.87 | 11.11 | 12.51 | 6.77 | 11.12 |
| Driving travel cost ($) | $Cost_{driving,t}$ | 0.31 | 0.25 | 0.23 | 0.32 | 0.25 | 0.24 |
| Biking travel time (min) | $TT_{biking,t}$ | 17.86 | 11.19 | 15.90 | 17.96 | 11.07 | 16.00 |
| Walking travel time (min) | $TT_{walking,t}$ | 40.98 | 25.13 | 38.75 | 41.07 | 24.83 | 39.05 |
| Carpool travel time (min) | $TT_{carpool,t}$ | 12.64 | 7.24 | 11.00 | 12.62 | 7.11 | 11.04 |
| Carpool travel cost ($) | $Cost_{carpool,t}$ | 0.16 | 0.12 | 0.12 | 0.16 | 0.12 | 0.12 |
| Microtransit in-vehicle time (min) | $TT_{MT,t}$ | 12.59 | 6.80 | 11.19 | 12.58 | 6.70 | 11.21 |
| Microtransit waiting time (min) | $WT_{MT,t}$ | 14.12 | 0.38 | 14.11 | 11.71 | 0.40 | 11.71 |
| Microtransit trip fare ($) | $COST_{MT,t}$ | 3.82 | 0.71 | 3.50 | 3.84 | 0.71 | 3.50 |
| Trip purpose = shopping | $SH_{n,t}$ | 0.17 | 0.38 | 0 | 0.27 | 0.42 | 0 |
| Trip purpose = school | $SC_{n,t}$ | 0.20 | 0.61 | 0 | 0.02 | 0.10 | 0 |
| Trip purpose = other | $OT_{n,t}$ | 0.64 | 0.48 | 1 | 0.71 | 0.46 | 1 |
| Tour type = commute | $COM_{n,t}$ | 0.45 | 0.50 | 0 | 0.13 | 0.34 | 0 |
| Tour type = non-commute | $NCOM_{n,t}$ | 0.55 | 0.52 | 1 | 0.87 | 0.36 | 1 |

Note: The unit of time is transformed to hours when estimating the model.

*4.1.4. Scenario Design*

Based on the choice model and simulation framework, we focus on two key revenue management scenarios, each addressing different aspects of pricing and subsidy strategies in the microtransit system.

In the ride-pass subscription pricing scenario, we aim to determine the optimal combination of weekly and monthly ride pass prices that maximizes total daily revenue. We explore how different pricing structures influence ridership levels and overall revenue generation, which is particularly relevant for optimizing subscription models, encouraging ridership, and enhancing the financial sustainability of the microtransit service. According to the current pricing policy, we set the range of weekly price to [$10, $40] and the range of monthly price to [$40, $100], each of them divided into 20 intervals, resulting in 20*20=400 pricing scenarios as inputs of our simulation framework. Metrics defined in Section 3.3.3 are used for evaluation.

In the event- or place-based subsidy scenario, we assess the impact of targeted discounts on microtransit trips linked to specific locations or events. These include essential infrastructure such as hospitals, where improved access can enhance healthcare equity, and high-traffic events like football games, where strategic subsidies could help mitigate congestion and reduce reliance on private vehicles. For the event-based subsidy, we focus on the events at AT&T Stadium, which is home of the Dallas Cowboys. For the place-based subsidy, we focus on Medical City Arlington, a major local healthcare facility. Three discounting scenarios are considered in our simulation: no discount, 50% discount, and 100% discount. For each scenario, we quantify the reduction in private car usage and estimate the additional subsidy required to maintain the affordability of these discounted fares.

## 4.2. Model results

This section presents the results of our behavior model. The experiments were conducted on a local machine equipped with an Intel Core i7-10875H CPU and 32GB of RAM. Optimization problems were solved using the Gurobi package in Python. Algorithm 2 took 17.15 hours and 3 iterations to converge for the nested model, with AMXL estimation in the lower branch accounting for the majority of the computing time. Algorithm 1 took 9.08 hours and 28 iterations to converge using weekday data containing 699,995 trips and 7.95 hours and 24 iterations to converge using weekend data containing 646,784 trips. Standard errors of the taste parameters are computed via bootstrapping with 30 resamples (Krueger et al., 2023). We present the model results from three aspects: (1) basic statistics; (2) taste parameter distribution; and (3) CS and elasticity estimates.

*4.2.1. Basic statistics*

Table 3 summarizes the values, standard errors, and significant levels of parameters in the upper-branch subscription choice model. All the estimated parameters are significant, and their signs align with our empirical expectations: (1) The positive parameters for benefits gained on weekdays ($\beta_{gain}^{we}$) and weekends ($\beta_{gain}^{wd}$) confirm that increased utility from microtransit usage encourages ride pass subscription; (2) The strong positive parameter for previous microtransit users ($\beta_{MT-user}$) reinforces the notion that prior usage significantly influences subscription likelihood; and (3) The negative constants for weekly ($asc_{weekly}$) and monthly ($asc_{monthly}$) passes reflect the low subscription rate (only 0.14% of the residents are subscribers), as Arlington Via has not yet been widely adopted.

**Table 3**
Results of the upper-branch subscription choice model

| Parameters | Mean | Std. Error | t-stat. |
|---|---|---|---|
| Ride pass price/trip cost ratio ($f_{rp}$) | 0.478 | 0.037 | 12.92*** |
| Benefits gained on weekdays ($\beta_{gain}^{we}$) | 3.156 | 0.422 | 7.48*** |
| Benefits gained on weekends ($\beta_{gain}^{wd}$) | 1.505 | 0.143 | 10.52*** |
| Previous microtransit user ($\beta_{MT-user}$) | 4.767 | 0.098 | 48.64*** |
| Constant for weekly pass ($asc_{weekly}$) | –0.844 | 0.128 | –6.59*** |
| Constant for monthly pass ($asc_{monthly}$) | –1.186 | 0.166 | –7.14*** |
| Total number of residents: 293,663 | | | |
| Observed weekly pass subscribers: 320 | Predicted monthly pass subscribers: 320 | | |
| Observed monthly pass subscribers: 92 | Predicted monthly pass subscribers: 93 | | |

The subscription model demonstrates high predictive accuracy. The predicted number of weekly pass subscribers (320) and monthly pass subscribers (93) closely match the observed values. Table 4 presents predicted number of subscribers across different demographic segments, which is unavailable without the estimated model. The not-low-income segment constitutes the majority of subscribers, accounting for 58.60% of the total. The student segment emerges as the second-largest subscriber group, comprising 28.81% of total subscribers. In contrast, the low-income segment exhibits a substantially lower subscription rate, representing only 2.42% of total subscribers, potentially due to financial constraints or limited service accessibility.

**Table 4**
Predicted ride-pass ownership by demographic segments

| | Weekly RP subscribers | Monthly RP subscribers | Total subscribers |
|---|---|---|---|
| Non-low-income segment | 188 (45.52%) | 54 (13.08%) | 242 (58.60%) |
| Low-income segment | 8 (1.94%) | 2 (0.48%) | 10 (2.42%) |
| Senior segment | 32 (7.75%) | 10 (2.42%) | 42 (10.17%) |
| Student segment | 92 (22.28%) | 27 (6.54%) | 119 (28.81%) |
| Total population | 320 (77.48%) | 93 (22.52%) | 413 (100%) |

Note: Each entry represents the number of subscribers, and the number in the parenthesis is the proportion.

Table 5 summarizes the mean values, standards error, and significant levels of parameters in the mode choice model for both weekday and weekend trips. All the estimated parameters are significant at 1% level. The models demonstrate a strong goodness-of-fit, as reflected by the McFadden R-square values (0.603 for weekdays and 0.576 for weekends). The mode-specific constants reveal a strong preference for private vehicle use in the study area, with a significantly positive constant for driving (1.433 on weekdays and 1.636 on weekends). In contrast, microtransit exhibits relatively low attractiveness as a travel mode, as evidenced by its negative mode-specific constant (–0.105 on weekdays and –0.121 on weekends) and negative interaction terms with various trip purposes and tour types. Additionally, a comparison between weekday and weekend models reveals that residents' sensitivity to mode attributes varies slightly across different days of the week. The relatively lower McFadden R-square in the weekend model suggests that travelers exhibit greater flexibility in their mode choices on weekends.

**Table 5**
Results of the lower-branch mode choice model

| Parameters | Weekday | | | Weekend | | |
|---|---|---|---|---|---|---|
| | Mean | Std. Error | T-stat. | Mean | Std. Error | T-stat. |
| **Travel time and cost** | | | | | | |
| Auto travel time ($\beta_{tt_auto,n}$) | –0.090 | 0.001 | –67.80*** | –0.081 | 0.001 | –69.26*** |
| Non-auto travel time ($\beta_{tt_non_auto,n}$) | –0.135 | 0.001 | –119.59*** | –0.127 | 0.001 | –121.03*** |
| Microtransit waiting time ($\beta_{wt,n}$) | –0.161 | 0.002 | –106.62*** | –0.161 | 0.001 | –132.82*** |
| Travel cost ($\beta_{cost,n}$) | –0.333 | 0.004 | –91.16*** | –0.307 | 0.003 | –93.47*** |
| **Mode specific constants** | | | | | | |
| Microtransit constant ($asc_{MT,t}$) | –0.105 | 0.001 | –85.97*** | –0.121 | 0.002 | –78.39*** |
| Driving constant ($asc_{driving,t}$) | 1.433 | 0.037 | 39.03*** | 1.636 | 0.032 | 51.79*** |
| Biking constant ($asc_{biking,t}$) | –0.712 | 0.005 | –131.06*** | –0.608 | 0.005 | –117.29*** |
| Walking constant ($asc_{walking,t}$) | 0.176 | 0.009 | 20.39*** | 0.125 | 0.008 | 16.34*** |
| **Trip purpose and tour type** | | | | | | |
| Trip purpose: shopping ($\beta_{shopping}$) | –0.012 | 0.000 | –47.85*** | –0.026 | 0.000 | –98.85*** |
| Trip purpose: school ($\beta_{school}$) | –0.035 | 0.000 | –134.37*** | –0.003 | 0.000 | –10.80*** |
| Trip purpose: other ($\beta_{other}$) | –0.043 | 0.000 | –164.12*** | –0.065 | 0.000 | –248.39*** |
| Tour type: commute ($\beta_{commute}$) | –0.038 | 0.000 | –147.46*** | –0.017 | 0.000 | –63.94*** |
| Tour type: non-commute ($\beta_{noncom}$) | –0.019 | 0.000 | –71.42*** | –0.103 | 0.000 | –396.01*** |
| **Meta information** | | | | | | |
| # Observations | 699,995 observations | | | 646,784 observations | | |
| Loglikelihood value (null) | –1,126,598 | | | –1,040,958 | | |
| Loglikelihood value (model) | –447,259 | | | –441,366 | | |
| McFadden R-square | 0.603 | | | 0.576 | | |

Significant levels (p-value): <0.01 '***' <0.05 '**' <0.1 '*'

Since travel time and cost parameters vary across individuals, we can further explore the value-of-time (VOT) of four population segments. The value of in-vehicle time is calculated by dividing the mean value of the auto travel time parameter ($\beta_{tt_auto,n}$) by the mean value of the travel cost parameter ($\beta_{cost,n}$). The value of waiting time is calculated by dividing the mean value of the microtransit waiting time parameter ($\beta_{wt,n}$) by the mean value of the travel cost parameter ($\beta_{cost,n}$). Table 6 lists the VOT by segments calculated using the weekday and weekend models. In general, the VOT on weekdays is higher than on weekends, and the value of microtransit waiting time is higher than in-vehicle time. Among the population segments, the non-low-income segment has the highest VOT while the low-income segment has the lowest VOT. These align with the findings in existing mode choice studies (Fournier & Christofa, 2021; Kolarova et al., 2018).

**Table 6**
Value of microtransit in-vehicle time and waiting time by demographic segments

| | Value of in-vehicle time ($/hour) | | Value of waiting time ($/hour) | |
|---|---|---|---|---|
| | Weekday | Weekend | Weekday | Weekend |
| Non-low-income segment | 17.36 | 15.22 | 28.61 | 25.98 |
| Low-income segment | 10.24 | 11.09 | 16.07 | 18.07 |
| Senior segment | 11.62 | 10.61 | 19.87 | 18.88 |
| Student segment | 14.10 | 11.10 | 22.47 | 18.19 |

*4.2.2. Distribution of taste parameters*

In our mode choice model, time and cost parameters vary across individuals, mode specific constants vary across trips, and interaction term parameters are homogeneous across all observations. Fig. 3 shows the statistical distribution of mode choice parameters, revealing them to be neither Gaussian nor other parametric distributions. Instead, these parameters can be divided into three categories: (1) evenly-distributed parameters, such as travel cost (B_COST) and driving constant (ASC_DRIVING). These parameters have larger variations, reflecting heterogeneous tastes across agents; (2) highly-concentrated parameters, such as microtransit constant (ASC_MICRO) and non-auto travel time (B_NON_AUTO_TT). These parameters are concentrated around their mean values with small variations, reflecting homogeneous tastes across agents; and (3) parameters following asymmetric (or log-normal) distributions, such as biking (ASC_BIKING) and walking (ASC_WALKING) constants. These parameters exhibit a peak value and a long tail in their distributions, indicating a small proportion of agents being more sensitive to an attribute compared to the majority. Additionally, although the parameters for trip purpose and tour type are assumed to be fixed, they exhibit greater differences between weekdays (Panel (c)) and weekends (Panel (f)) compared to other parameters.

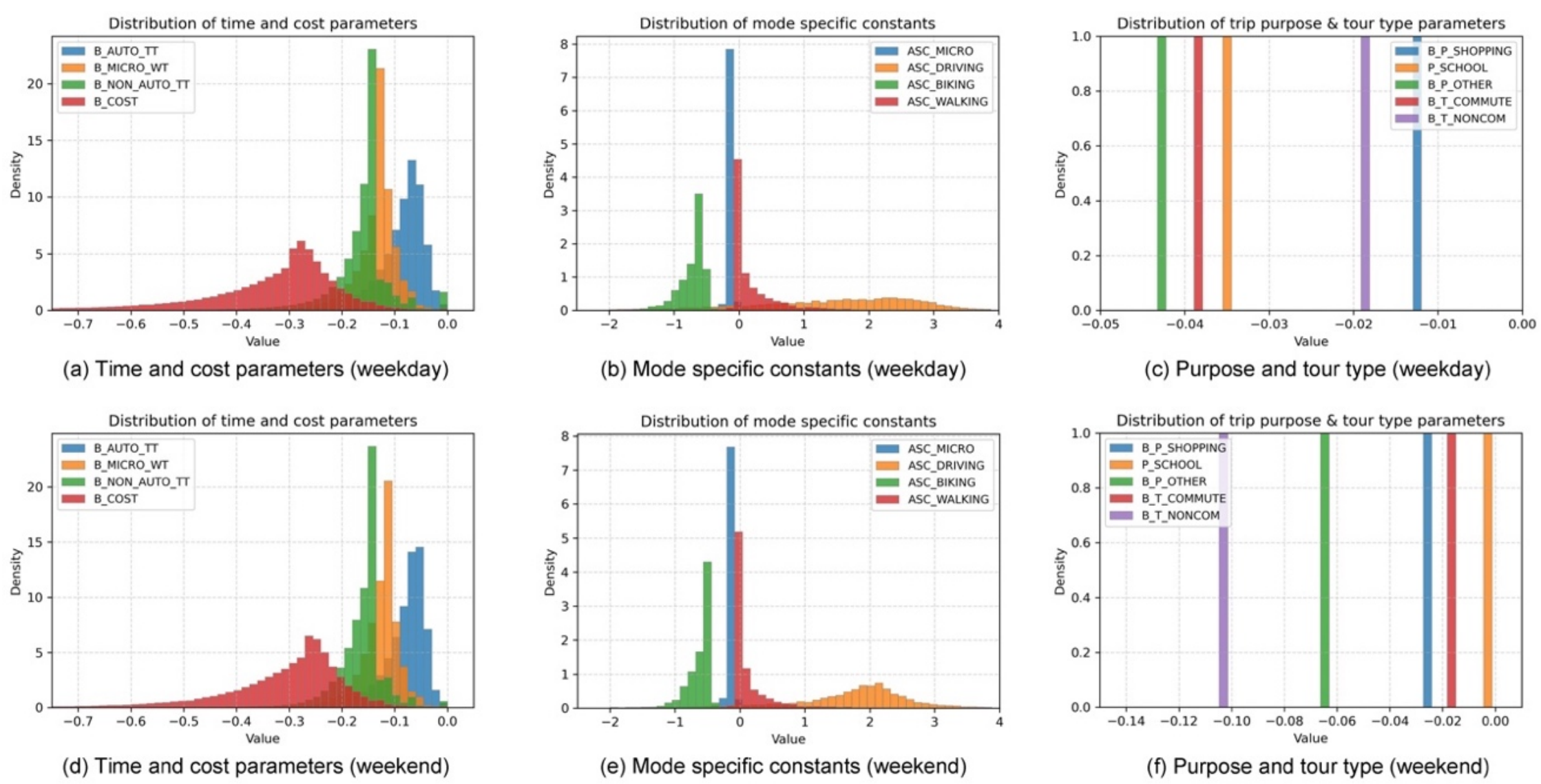

**Fig. 3.** Statistical distributions of mode choice parameters

Fig. 4 presents the spatial distribution of constants for driving, walking, and microtransit, alongside Arlington's zoning districts (City of Arlington, 2022). Driving constants are generally higher along major roadways and corridors, suggesting a strong preference for driving in well-connected areas. Walking constants exhibit localized variations, with higher values concentrated in denser urban areas, likely reflecting pedestrian-friendly environment. Microtransit constants are generally lower and evenly distributed, probably because microtransit is not widely adopted. The spatial distribution of these constants aligns with land use patterns in Panel (g), where commercial and mixed-use areas support higher walking activity, while lower-density residential zones are more car-dependent. The results highlight the capability of our mode choice models in capturing the influence of the built environment on user preferences.

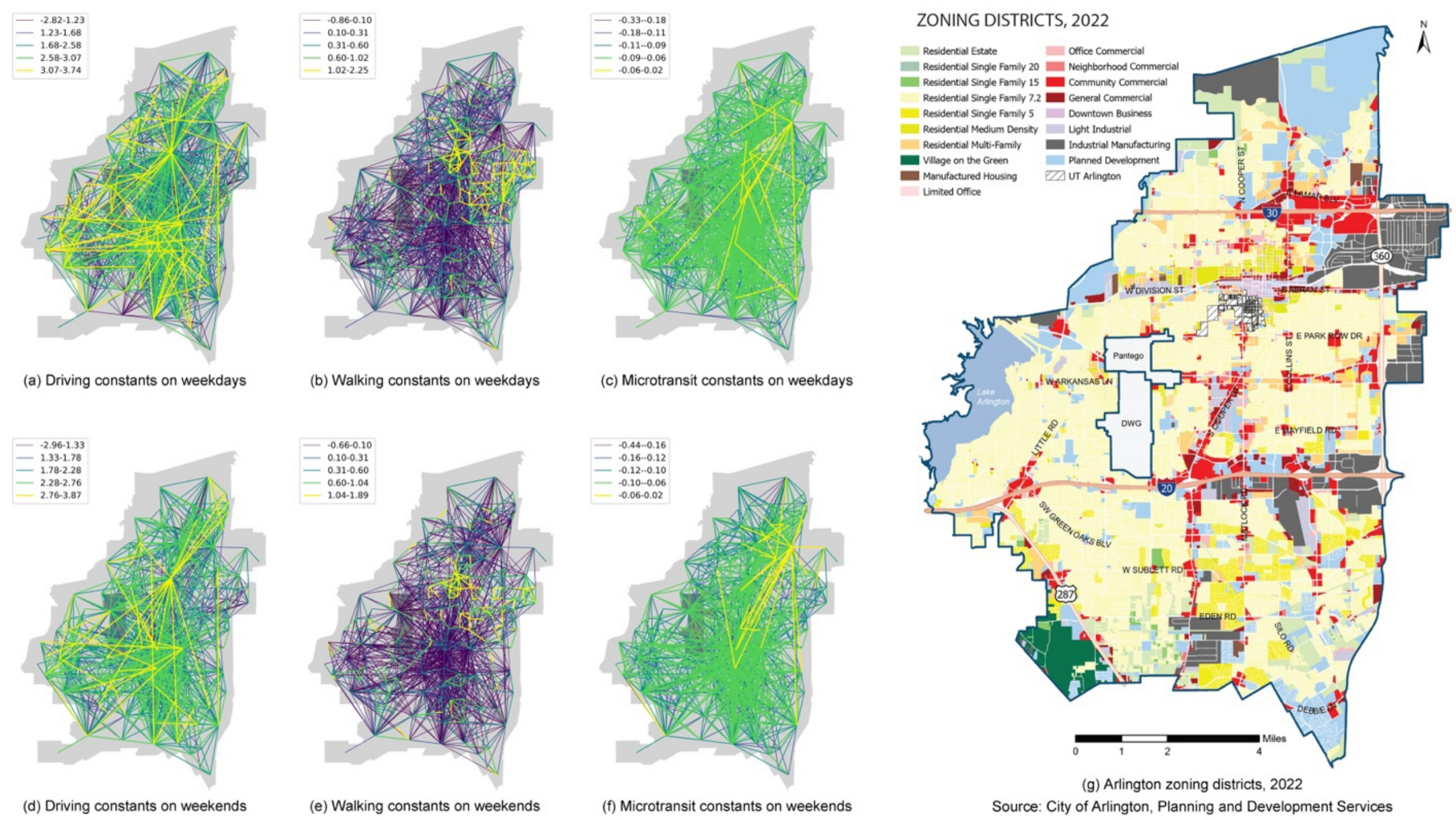

**Fig. 4.** Spatial distributions of mode specific constants

*4.2.3. Consumer surplus and elasticity estimates*

Fig. 5 illustrates the distribution of subscription-based and trip-level compensating variation (CV) across four population segments, highlighting differences in how they value ride pass plans and the microtransit service. Panel (a) depicts the CV for having both weekly and monthly ride pass options, ranging from approximately \$0 to \$0.20/week. Notably, the not-low-income segment exhibits the highest values, with a distribution peak around \$0.03/week. In contrast, the senior and low-income segments show relatively lower CV values, with most values concentrated near \$0.01/week. The relatively low value is probably because microtransit occupies such a small mode share.

Panels (b) and (c) compare microtransit CV on weekdays and weekends. In general, weekday CV distributions exhibit greater dispersion across individuals and segments, implying more diverse attitudes toward microtransit on weekdays compared to weekends. The density functions indicate that the low-income and senior segments have the most concentrated peaks near \$0.01/week, suggesting that these groups experience relatively consistent but modest perceived value. The not-low-income segment shows a broader spread, with peaks occurring at slightly higher values, implying higher perceived value and a wider range of user experiences.

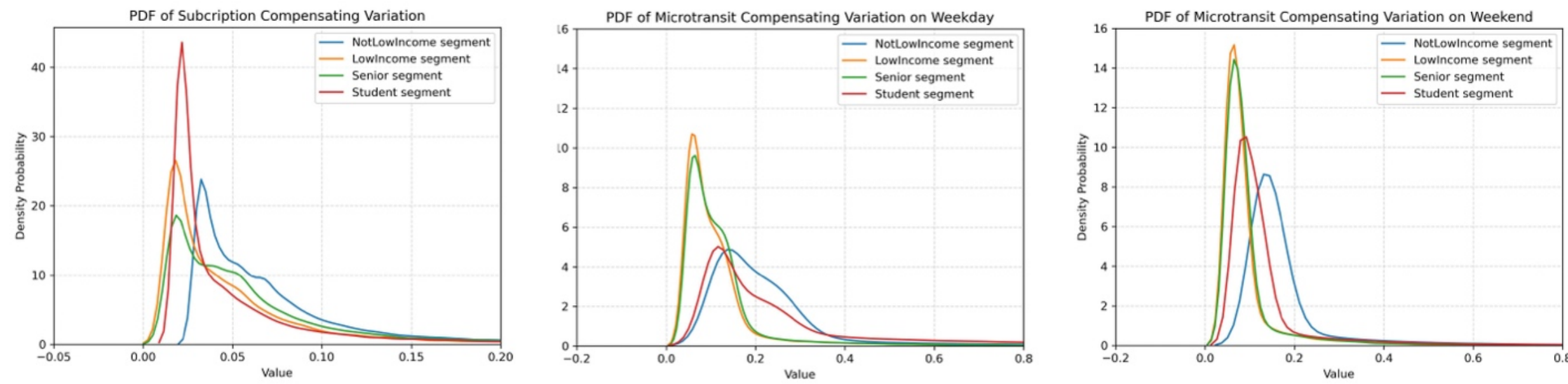

(a) PDF of ride pass CV (b) PDF of weekday microtransit CV (c) PDF of weekend microtransit CV

**Fig. 5.** Probability distribution functions (PDF) of compensating variation (CV) by population segments

Fig. 6 presents the cumulative distribution functions (CDFs) of ride pass elasticity with respect to microtransit service performance on weekdays (Panel (a)) and weekends (Panel (b)). The results reveal that elasticities are consistently negative, indicating that improvements in microtransit service performance increase the likelihood of ride pass subscription. However, the magnitude of elasticity is relatively small, indicating that the relationship between ride pass ownership and service performance falls within the inelastic region, likely due to the small proportion of microtransit users at this stage. Furthermore, elasticity values are more negative on weekdays than on weekends and more negative with respect to waiting time than in-vehicle time, suggesting that quicker service responses during weekdays have a stronger impact on ride pass subscription. These interpretable results not only validate the value of our model estimation but also provide valuable insights for designing targeted pricing and subsidy strategies.

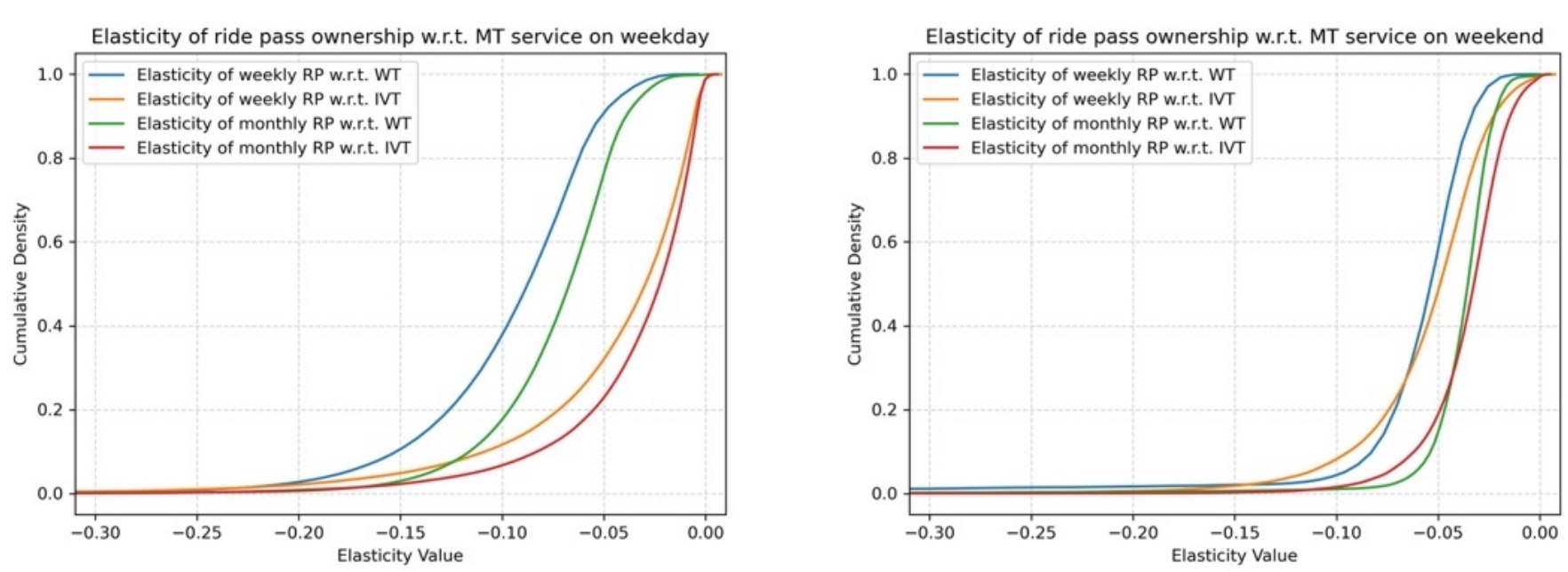

(a) CDF of elasticities w.r.t. service performace on weekdays (b) CDF of elasticities w.r.t. service performace on weekends

**Fig. 6.** Cumulative distribution function (CDF) of ridepass ownership elasticity w.r.t. microtransit service performance

*4.2.4. Sensitivity test*

As reported in Section 4.1.2, the simulator produces approximately 15% prediction error for the Arlington case study when benchmarked against observed ridership data. Given that the current microtransit market share in the study area is very small (less than 0.5%), a natural concern arises: could even moderate simulation errors be amplified as they propagate through the downstream mode choice and ride-pass subscription predictions, leading to disproportionately large distortions in the key planning metrics? To address this concern, we conduct a sensitivity analysis examining how the three key outputs—total ridership (trips/day), total ride-pass subscribers (persons/week), and total revenue (\$/day)—respond to deviations in microtransit service performance across different market share scenarios.

The sensitivity test is designed along two dimensions. First, to assess the impact of microtransit simulation errors, we perturb the microtransit in-vehicle time (IVT, the variable exhibiting the largest discrepancy) by applying percentage changes ranging from −50% to +50% in 5% increments to the baseline values in the choice dataset. This range is deliberately wide to test the model predictions well beyond the observed simulation error. It is worth noting that this approach assumes errors in microtransit in-vehicle time do not significantly affect the estimated parameters on auto travel time ($\beta_{auto_tt,n}$), which is reasonable given that this parameter also depends on driving and carpool travel times recorded in Replica's trip data. Second, we examine these IVT perturbations under five market share scenarios: the current level (<0.5%), and hypothetical levels of approximately 5%, 10%, 20%, and 40%. The higher market share scenarios are achieved by adjusting the alternative-specific constant for microtransit ($asc_{MT,t}$) in the mode choice model. Since a higher microtransit market share also increases the upper-branch utility of having a ride pass, we simultaneously update the ride-pass subscription predictions at each scenario to maintain a consistent nested feedback loop.

Fig. 7 presents the percentage deviations of the three metrics from their respective baseline values as the microtransit IVT varies from −50% to +50%. At the current market share, all three metrics remain relatively flat across the entire perturbation range, with deviations staying within ±5% given a ±15% perturbation. The 20% scenario exhibits the largest deviations, up to roughly 60% for subscribers under the –50% perturbation of IVT. Across all scenarios, overestimation of in-vehicle time leads to underestimation of ridership, subscribers, and revenue, and vice versa. The magnitude of this bias, however, depends on the market share scenario.

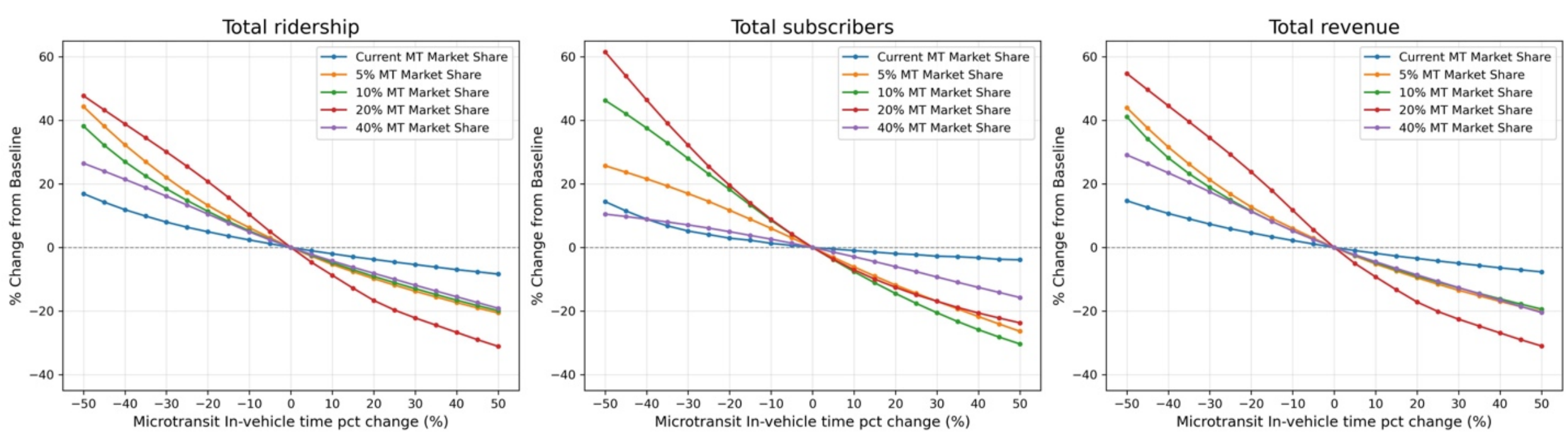

**Fig. 7.** Pct deviations of key metrics w.r.t. microtransit IVT under five market share scenarios

Fig. 8 provides deeper insight by plotting the elasticity of each metric with respect to in-vehicle time, computed as the ratio of the percentage deviations in the metric from its baseline value to the corresponding percentage change in IVT. At the current market share, all three elasticities are around –0.2, indicating that the predictions are inelastic to IVT perturbations. Such robustness can be intuitively attributed to the logistic (sigmoid) structure of the logit model: when a mode's market share is very small, the choice probability sits on the flat tail of the sigmoid curve, where even substantial changes in the utility argument produce only marginal shifts in the predicted probability. A detailed explanation is provided in Appendix A4. In practical terms, this means that under early-stage microtransit adoption, simulation errors in IVT are dampened rather than amplified as they propagate through the mode choice and subscription models.

The 20% scenario is notably the most elastic across all three metrics, with some values falling in the elasticity region (elasticity magnitude larger than 1.0). This is also expected given the S-shaped probability response inherent to the logit model—as microtransit's utility improves, the mode's choice probability transitions through the steep, sensitive middle portion of the sigmoid curve. The 40% scenario, by contrast, shows lower elasticities than the 20% case, likely because its baseline probability has already moved into the flatter upper portion of the sigmoid.

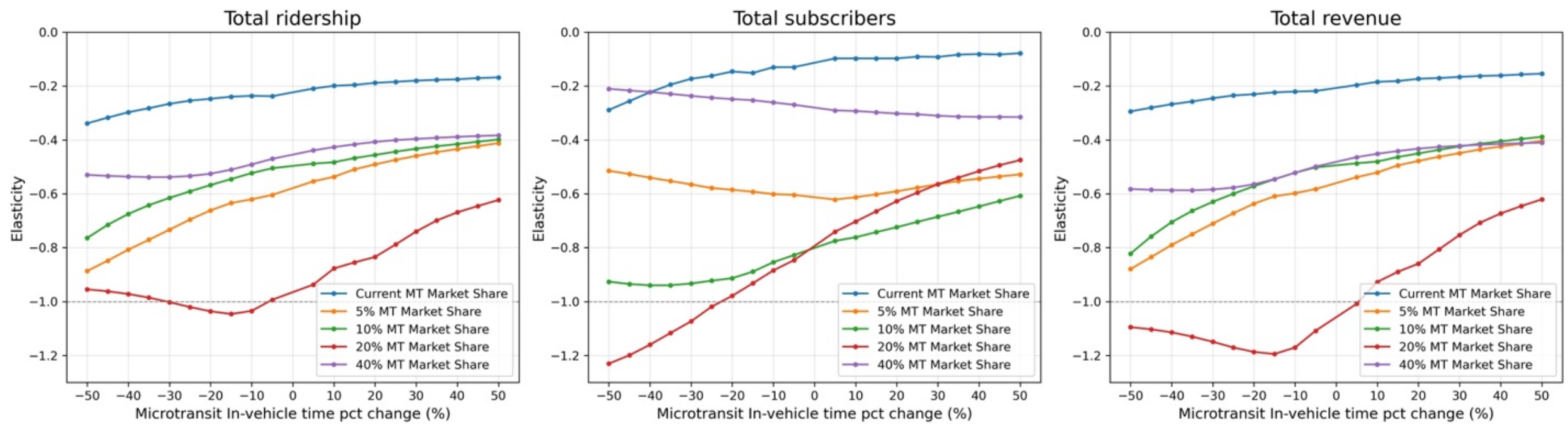

**Fig. 8.** Elasticity of key metrics w.r.t. microtransit IVT under five market share scenarios

These findings carry two important implications for the integrated framework. First, they provide evidence that, at the current early stage of microtransit deployment, the simulation errors in our case are not amplified during the downstream propagation. The inelasticity portion of the logit structure reduces the impact of IVT noise on model predictions. Second, the results highlight that as microtransit matures and its market share grows toward 20%, the system's sensitivity to simulation accuracy would increase substantially. In this transition phase, the fidelity of microtransit simulation becomes increasingly important for ridership, subscription, and revenue forecasts.

## 4.3. Policy evaluation for revenue management

This section demonstrates how our estimated model can be integrated into a simulation-based framework to support revenue management policy evaluation. The simulations were conducted on a local machine equipped with an Intel Core i7-10875H CPU and 32GB of RAM, with computing times per simulation ranging from 1 to 3 minutes.

### *4.3.1. Ride-pass subscription pricing*

We evaluate the impact of adjusting ride pass prices on total daily revenue by simulating 400 pricing scenarios. Fig. 9 presents how total daily revenue, number of subscribers, and microtransit ridership change based on weekly and monthly ride pass prices. The results show that the optimal pricing strategy to maximize total daily revenue is to reduce the weekly ride pass price from $25 to $18.9 and the monthly ride pass price from $80 to $71.5. This pricing adjustment increases total daily revenue from $11,956 to $12,083, a net gain of $127 per day. Panel (b)-(c) further show why the total revenue increase with the decrease of ride pass price. The revenue growth can be attributed to a higher number of subscribers (635 compared to 413 at the current pricing) and an increase in total microtransit ridership (3,225 trips compared to 3,105 trips). Lowering the price makes the ride pass more attractive to potential subscribers, expanding the customer base and boosting

overall ridership, which in turn compensates for the reduced per-pass price and results in higher total revenue.

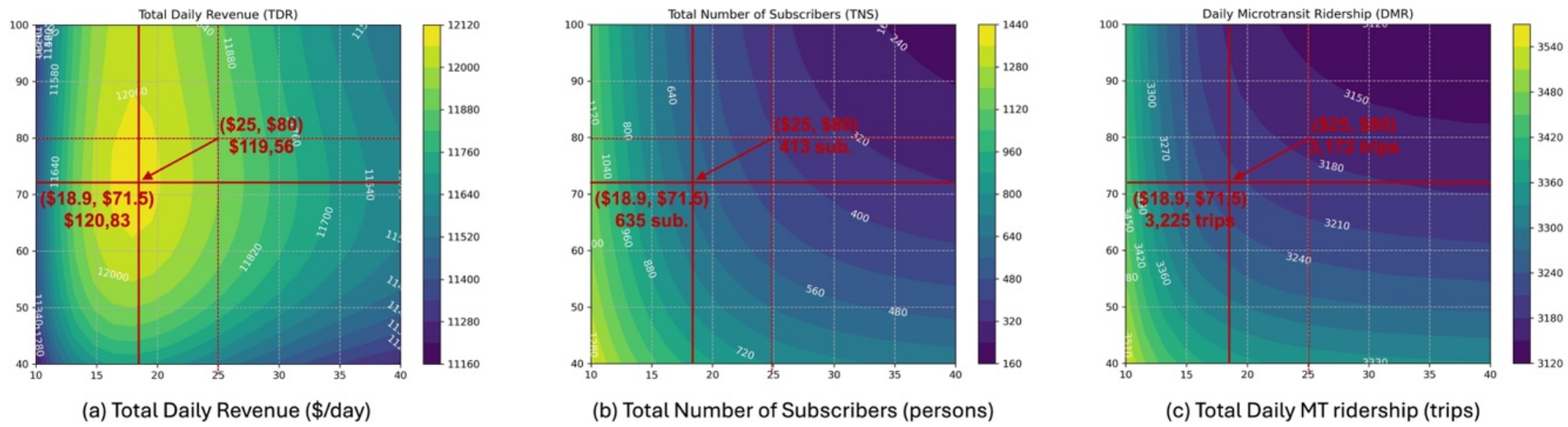

**Fig. 9.** Contour plots of total daily revenue (a), number of subscribers (b), and microtransit ridership (c) as a function of ride pass prices

### *4.3.2. Event-based subsidy policy for AT&T Stadium*

For the event-based subsidy policy, we focus on the impact of fare discounts on microtransit trips to and from AT&T Stadium, home of the Dallas Cowboys (Fig. 10). The analysis focuses on peak hours from  4 p.m. to 8 p.m. on weekends, coinciding with major sporting events.

The simulation results indicate that implementing a 100% fare discount for trips starting from or ending at AT&T Stadium leads to an increase in microtransit ridership from 78 to 141 trips per event. This increase helps shift some trips away from private vehicle usage, reducing the number of driving trips from 2,207 trips per event to 2,146 trips per event, which corresponds to a reduction of 61 car trips during peak periods. Additionally, the CS per trip increases by \$0.018 due to the fare discount, translating into a total monetary value of \$0.018 × 3,559 = \$ 64.62 across all affected trips. However, this policy would require a subsidy of \$533 per event to compensate the revenue loss. A 50% fare discount yields a more moderate impact, increasing microtransit ridership to 97 trips per event, reducing 18 car trips during peak period, while requiring a lower subsidy of \$193 per event. Although the simulated policy impact is relatively small compared to the total trip volume (largely due to the currently low mode share of microtransit), our model and simulation framework provide a quantitative measurement tool for evaluating such policies

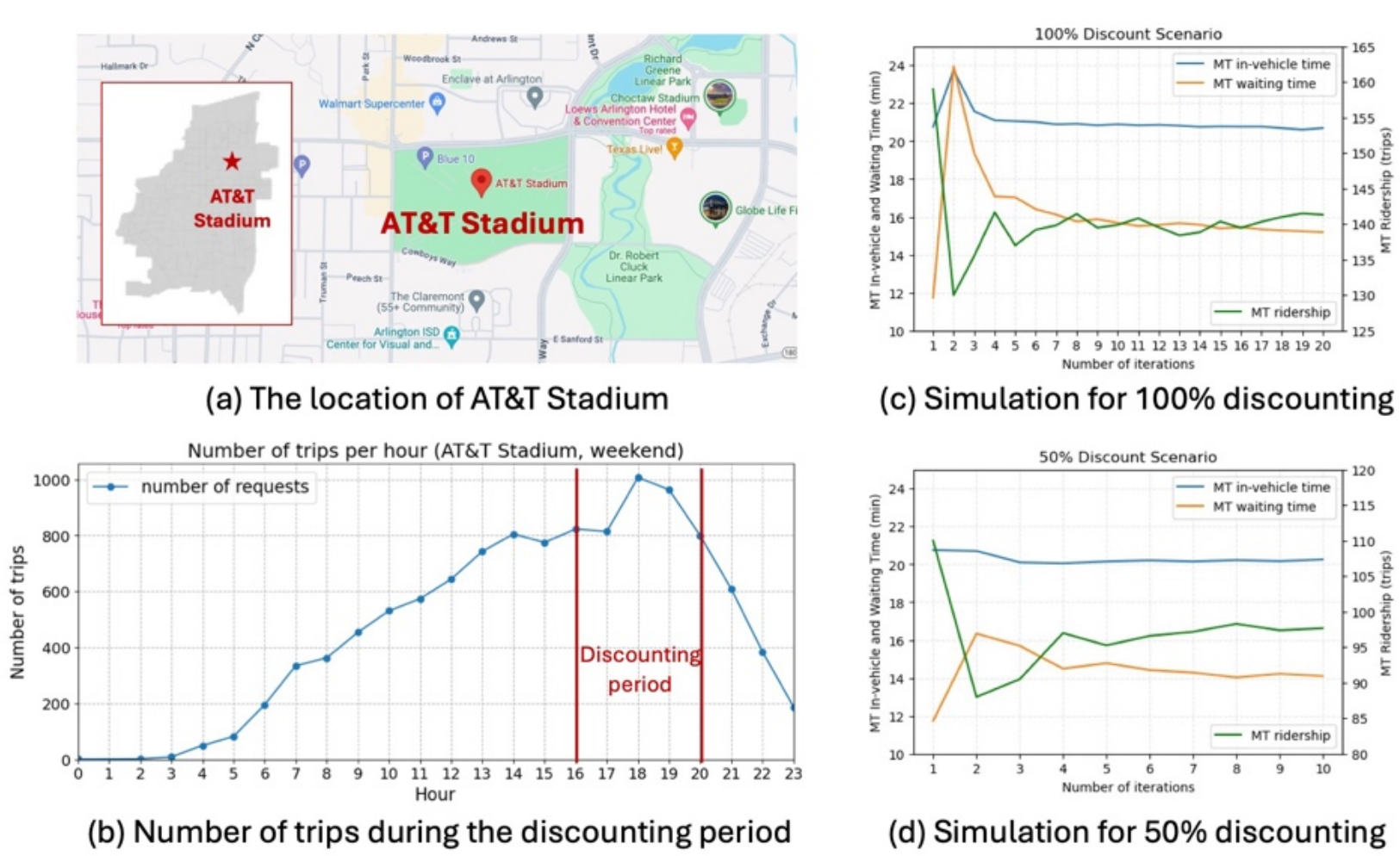

| | 100% discount | 50% discount | No discount |
|---|---|---|---|
| Total trips impacted | 3,599 | 3,599 | 3,599 |
| Microtransit trips | 141 | 97 | 78 |
| Driving trips | 2,146 | 2,189 | 2,207 |
| Microtransit in-vehicle time | 29.66 min | 27.92 min | 27.67 min |
| Microtransit waiting time | 14.31 min | 13.13 min | 11.73 min |
| Increased trip CS | \$0.018 per trip | \$0.009 per trip | \$0 per trip |
| Revenue loss | \$533 per event | \$193 per event | \$0 per event |

(e) Summary of simulation results

**Fig. 10.** A summary of two discounting scenarios for trips to AT&T stadium

*4.3.3. Place-based subsidy policy for Medical City Arlington*

For the place-based subsidy policy, we analyze the impact of fare discounts on microtransit trips to and from Medical City Arlington, a key healthcare facility in the region (Fig. 11). The analysis focuses on weekdays, particularly during the discounting period from 6 a.m. to 9 p.m., when a significant number of trips are made to the medical center. The goal of this policy is to improve access to healthcare facilities by encouraging microtransit usage through fare reductions.

The simulation results indicate that implementing a 100% fare discount for trips starting from or ending at Medical City Arlington increases microtransit ridership from 42 to 124 trips per day, which corresponds to 82 more microtransit trips during daytime. Additionally, the CS per trip increases by \$0.025, translating into a total monetary value of \$0.025 × 6,322 = \$158.05 across all affected trips. However, this policy would require a subsidy of \$483 per day to compensate for the revenue loss. A 50% fare discount yields a more moderate impact, increasing microtransit ridership to 67 trips per day, reducing 24 car trips, while requiring a lower subsidy of \$145 per day.

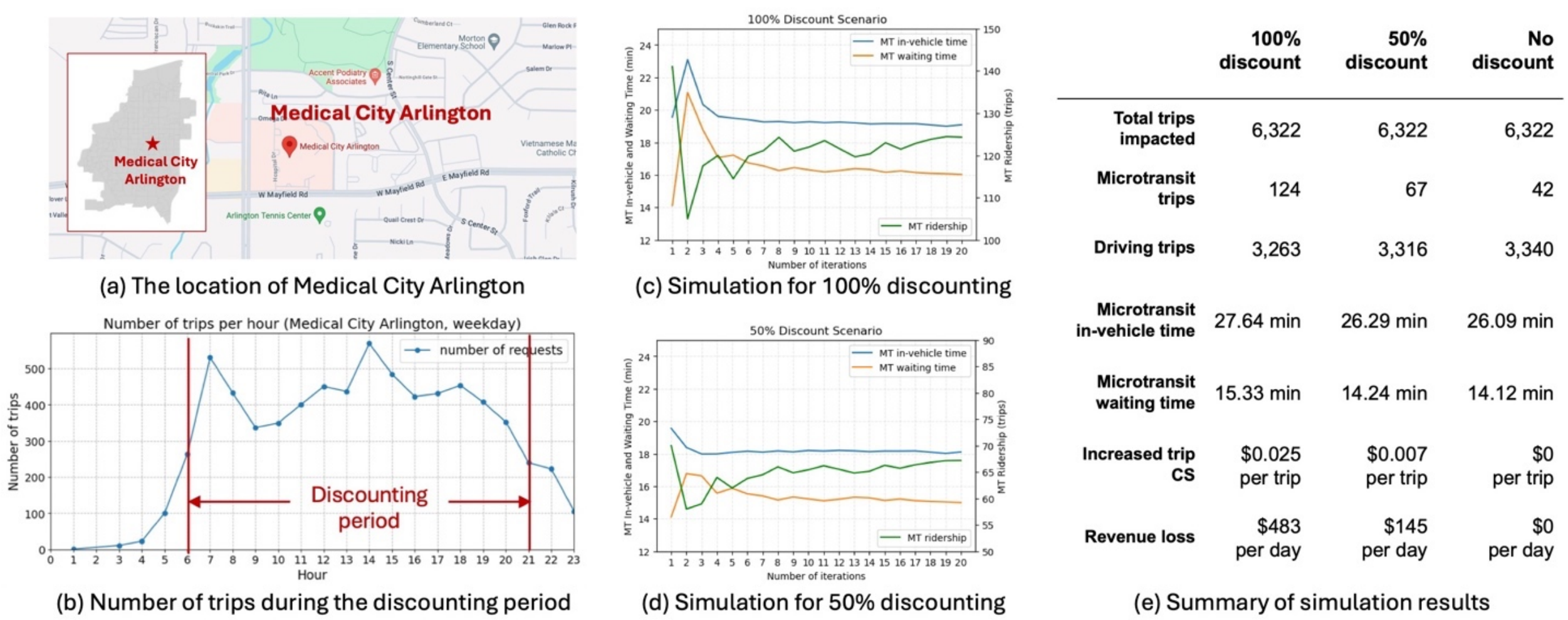

| | 100% discount | 50% discount | No discount |
|---|---|---|---|
| Total trips impacted | 6,322 | 6,322 | 6,322 |
| Microtransit trips | 124 | 67 | 42 |
| Driving trips | 3,263 | 3,316 | 3,340 |
| Microtransit in-vehicle time | 27.64 min | 26.29 min | 26.09 min |
| Microtransit waiting time | 15.33 min | 14.24 min | 14.12 min |
| Increased trip CS | \$0.025 per trip | \$0.007 per trip | \$0 per trip |
| Revenue loss | \$483 per day | \$145 per day | \$0 per day |

**Fig. 11.** A summary of two discounting scenarios for trips to Medical City Arlington

# 5. Discussion

## 5.1. Assumptions and limitations

Our study rests on several simplifying assumptions that merit explicit acknowledgment. First, due to data availability constraints, the upper-branch model is calibrated using only aggregate data rather than individual-level revealed or stated preference data. As a result, the model assumes homogeneous taste parameters across the population, and standard errors are obtained through bootstrapping rather than from maximum likelihood estimation on disaggregate choice observations (Train, 2009). While this approach is a practical necessity given the nascent stage of microtransit subscription programs, it does limit the model's ability and challenge its stability. As microtransit data become more widely available, it will become feasible to estimate more behaviorally rich upper-branch models with heterogeneous specifications and explicit uncertainty quantification.

Second, the microtransit simulator produces approximately 15% prediction error in in-vehicle travel time when benchmarked against observed data from the City of Arlington. This discrepancy may stem from two sources: the coarse grid-based calibration procedure used to set the simulator's operational parameters, and the fact that the simulation tool was not developed for the specific street network and demand patterns of Arlington. Both factors are amenable to future improvement through finer-grained calibration and context-specific model tuning. Fortunately, the sensitivity analysis presented in Section 4.2.4 demonstrates that, at the current microtransit market share (<0.5%), this level of simulation error does not meaningfully propagate through the downstream mode choice and ride-pass subscription models.

Third, the empirical analysis is based on a single case study in Arlington, TX—a context representative of low-density, car-oriented cities where microtransit is increasingly being deployed as an alternative to traditional fixed-route service (Ghimire et al., 2024; Volinski et al., 2019). The evaluated policy scenarios reflect the pricing structures currently under consideration by Arlington's transit authority. In addition, some operational dimensions are simplified: fleet size is held fixed, fare prices are static, and the service zone is uniform. The present study focuses on the most common pricing strategies observed in microtransit deployments. However, to our knowledge, no prior study has integrated a microtransit simulator with a joint subscription–mode choice model for revenue management, and the present work represents an initial test of this integrated framework. Moreover, these simplifications are not inherent limitations of the framework itself but rather reflect the scope of the current application. The approach is readily extensible to richer policy scenarios, including dynamic pricing, variable fleet sizing, zone-based fare structures, and time-of-day differentiation (Chin et al., 2016).

### 5.2. Policy implications for microtransit development

While the case study is situated in Arlington, the findings carry broader relevance for the growing number of microtransit deployments across the United States, including RIDE in Wilson, NC, Metro Flex in Omaha, NE, and SAM On Demand in Sioux Falls, SD. These services share key structural features with the Arlington context (low-density service areas, small vehicle fleets, and app-based booking), making the insights from our analysis transferable. Drawing on the empirical results, we highlight three broader policy implications.

First, a win-win pricing opportunity may exist at current adoption levels. Our evaluation results indicate that a modest reduction in ride-pass pricing can simultaneously increase subscriber uptake and improve total revenue. This finding may appear counterintuitive but reflects the elastic demand response at the current price point. When the cost of a subscription falls, the marginal rider attracted to the pass generates subscription revenue that more than offsets the foregone per-ride fare income. This suggests that microtransit operators have an opportunity to optimize their pricing policies to better balance revenue from subscriptions and out-of-pocket fares. Given the financial pressures facing many microtransit programs (Pinski et al., 2026), identifying such win-win pricing configurations is of considerable practical value.

Second, revenue management must account for demand-supply feedback effects. While fare discounts and subscription passes can attract additional riders, increased demand may degrade service performance through longer wait times and in-vehicle travel times, potentially pushing some users back toward competing modes. This generates a rebound effect analogous to those documented in road transport (Hymel et al., 2010). The dynamic adjustment process in our framework provides a step toward capturing these interdependencies, revealing clear rebound effects after event- and place-based fare discounts (Panels (c)–(d) in Figs. 11–12). To this end, effective revenue management for microtransit requires an integrated perspective that jointly considers the equilibrium between demand and supply.

Last but not least, the importance of quantitative revenue management tends to grow with microtransit market share. The sensitivity analysis reveals that key planning metrics are relatively inelastic to changes in microtransit service attributes at current stage. However, as microtransit matures and its market share grows toward 10–20%, the system transitions into an increasingly elastic regime where pricing, service quality, and simulation fidelity all exert substantially larger effects on ridership, subscriptions, and revenue. This trajectory suggests that while informal or heuristic approaches to fare-setting may suffice in the early stages, the growing sensitivity of outcomes to policy parameters will make quantitative decision-support models and higher quality data to support them progressively more valuable as microtransit services continue to scale (Mayaud et al., 2021). For agencies currently operating or planning microtransit deployments, investing in modeling capacity can yield significant returns as the service matures and the flexibility of operational strategies increase.

## 6. Conclusion

As microtransit services continue to evolve, understanding their adoption and revenue management remains a critical challenge. This study introduces a nested nonparametric logit model that jointly estimates travel mode and ride pass subscription choices using a combination of marginal subscription and synthetic population data. By leveraging agent-based estimation, the model captures heterogeneous user preferences while maintaining theoretical consistency with discrete choice modeling. Our choice-based simulation framework further extends this capability, enabling policymakers to assess pricing and subsidy policies with quantitative precision.

The methodological contributions of this work are threefold. First, the integration of synthetic population data (1.35 million trips made by 0.4 million Arlington residents) enables citywide demand estimation with spatiotemporal granularity, overcoming the limitations of small-sample SP surveys. Second, the agent-based mixed logit (AMXL) approach captures individualized preferences without restrictive parametric assumptions, revealing asymmetric and multimodal distributions of taste parameters that conventional mixed logit models fail to represent as well as achieving high McFadden $R^2$ values (0.603 for the weekday model and 0.576 for the weekend model). Third, the nested structure bridges subscription and mode choice dynamics, enabling operators to quantify subscriber distribution across demographic segments, with 58.60% from the not-low-income segment, 2.42% from the low-income segment, 10.17% from the senior segment, and 28.81% from the student segment.

The policy evaluation highlights key insights into microtransit pricing and subsidy strategies. Our simulation results suggest that lowering ride pass prices—from $25 to $18.9 for weekly passes and $80 to $71.5 for monthly passes—can increase total revenue by $127 per day by attracting more subscribers and riders. Event- and place-based fare discounts at AT&T Stadium and Medical City Arlington can encourage microtransit use, but require subsidies of $533 per event and $483 per day, respectively. The sensitivity analysis further shows that key planning metrics remain inelastic to simulation errors in microtransit in-vehicle time at the current stage, while as market share grows, the system transitions into an increasingly elastic regime where pricing decisions and simulation fidelity exert substantially larger effects on predicted outcomes.

Several promising directions emerge for future research. First, while this study evaluates static pricing policies with a fixed fleet size, the framework is readily extensible to richer policy scenarios, including dynamic pricing, variable fleet sizing, and zone-based fare structures. Second, extending the model to multi-modal networks and incorporating service-level optimization could yield deeper insights into microtransit's role in integrated urban mobility systems. Third, with the increasing availability of subscription data across microtransit deployments, it becomes feasible to estimate more behaviorally rich upper-branch models with confidence intervals and explicit uncertainty quantification. By refining these aspects, the proposed framework can support more data-driven decision-making in microtransit planning and operations across a wider range of cities and transit contexts.

## Acknowledgements

The authors were partially supported by the C2SMARTER Center (Award #69A3551747124). Data shared by Replica Inc. and City of Arlington (CoA) are gratefully acknowledged.

## CRediT authorship contribution statement

The authors confirm contribution to the paper as follows: study conception and design: Joseph Y. J. Chow, Xiyuan Ren, Venktesh Pandey; data collection: Joseph Y. J. Chow, Xiyuan Ren; analysis and interpretation of results: Xiyuan Ren, Joseph Y. J. Chow, Linfei Yuan; draft manuscript preparation: Xiyuan Ren, Joseph Y. J. Chow, Venktesh Pandey. All authors reviewed the results and approved the final version of the manuscript.

## Declaration of generative AI and AI-assisted technologies in the writing process

During the preparation of this work the authors used ChatGPT and Claude in order to improve readability and language. After using this tool/service, the authors reviewed and edited the content as needed and take full responsibility for the content of the published article.

## Declaration of competing interest

The authors declare no competing interests.

# Appendix

## A1. Notations in the choice model

**Table A1**
Notations used in the proposed model

| | |
|---|---|
| $N$ | A set of individuals indexed by individual ids |
| $T$ | A set of trips indexed by trip origin-destination (OD) pair |
| $S$ | A set of subscription alternatives |
| $J$ | A set of mode alternatives including microtransit |
| $P$ | A set of trip purpose |
| $Q$ | A set of tour type |
| $U_{n,weekly}$, $\varepsilon_{n,weekly}$ | Total and random utilities of individual $n$ purchasing weekly ride pass |
| $U_{n,monthly}$, $\varepsilon_{n,monthly}$ | Total and random utilities of individual $n$ purchasing monthly ride pass |
| $U_{n,none}$, $\varepsilon_{n,none}$ | Total and random utilities of individual $n$ not purchasing any ride pass |
| $U_{n,MT,t}$, $\varepsilon_{n,MT,t}$ | Total and random utilities of individual $n$ choosing microtransit in trip $t$ |
| $U_{n,j,t}$, $\varepsilon_{n,j,t}$ | Total and random utilities of individual $n$ choosing mode $j$ in trip $t$ |
| $V_{n,s}$ | Systematic utility if individual $n$ choosing ride pass option $s$ |
| $V_{n,j,t}$ | Systematic utility of individual $n$ choosing mode $j$ in trip $t$ |
| $V_{n,j,t}^{rp}$, $V_{n,j,t}^{nrp}$ | Systematic utility of individual $n$ choosing mode $j$ in trip $t$ with/without a ride pass |
| $Price_{weekly}$, $Price_{monthly}$ | Prices of weekly and monthly ride pass |
| $Gain_n^{we}$, $Gain_n^{wd}$ | "Benefits" gained from holding a ride pass on weekdays/weekends |
| $MT_n$ | A binary variable indicating if individual $n$ is a microtransit user |
| $TT_{MT,t}$, $TT_{j,t}$ | Travel time of microtransit/mode $j$ in trip $t$ |
| $WT_{MT,t}$ | Waiting time of microtransit in trip $t$ |
| $COST_{n,MT,t}$ | Microtransit fare (distance-based) charged to individual $n$ for trip $t$ |
| $COST_{j,t}$ | Fare of other mode $j$ in trip $t$ |
| $SH_{n,t}$ | A binary variable indicating if individual $n$'s trip $t$ is for shopping |
| $SC_{n,t}$ | A binary variable indicating if individual $n$'s trip $t$ is to school |
| $OT_{n,t}$ | A binary variable indicating if individual $n$'s trip $t$ serves other purposes |
| $COM_{n,t}$ | A binary variable indicating if individual $n$'s trip $t$ is part of a commute tour |
| $NCOM_{n,t}$ | A binary variable indicating if individual $n$'s trip $t$ is part of a non-commute tour |
| $\beta_{price,n}$ | Individual-level parameter of ride pass price |
| $\beta_{tt,n}$ | Individual-level parameter of travel time |
| $\beta_{wt,n}$ | Individual-level parameter of microtransit waiting time |
| $\beta_{cost,n}$ | Individual-level parameter of travel cost |
| $asc_{MT,t}$, $asc_{j,t}$ | Trip OD-level mode-specific constants |
| $\beta_{shopping}$ | Generic parameter of trips for shopping |
| $\beta_{school}$ | Generic parameter of trips to school |
| $\beta_{other}$ | Generic parameter of trips serving other purpose |
| $\beta_{commute}$ | Generic parameter of trips in commute tours |
| $\beta_{noncom}$ | Generic parameter of trips in non-commute tours |
| $\beta_{gain}^{we}$, $\beta_{gain}^{wd}$ | Generic parameter of "benefits" gained on weekdays/weekends |
| $\beta_{MT-user}$ | Generic parameter of microtransit user |
| $asc_{weekly}$, $asc_{monthly}$ | Generic ride pass-specific constants |
| $f_{rp}$ | Generic factor linking the parameters of travel cost and ride pass price |

## A2. Estimation algorithm for the choice model

Following the work of Ren and Chow (2022), we employ an agent-based mixed logit (AMXL) approach to estimate the lower-branch mode choice model for each individual and trip. In AMXL, each trip $t$ made by individual $n$ is represented as an agent, whose parameters ($\beta_{n,t}$) can be jointly and nonparametrically estimated by solving a multiagent inverse utility maximization (MIUM) problem under $L_2$-norm as a convex quadratic programming (QP) problem. We extend the approach of Ren and Chow (2022) by introducing additional constraints to ensure the consistency of parameters among individuals and trips, thereby enhancing model flexibility by allowing parameters to vary at different levels. The formulation of the estimation problem is shown in Eqs. (A1) – (A7).

$$\min_{\beta_0,\beta_{n,t}} \sum_{n\in N}\sum_{t\in T}(\beta_0-\beta_{n,t})^2 \tag{A1}$$

subject to:

$$V_{n,j^*,t}+\varepsilon_{n,j^*,t} \geq V_{n,j,t}+\varepsilon_{n,j,t}+b, \qquad \forall n\in N, t\in T, j, j^*\in J, j\neq j^* \tag{A2}$$

$$\beta_{n,t} \geq lb, \qquad \forall n\in N, t\in T \tag{A3}$$

$$\beta_{n,t} \leq ub, \qquad \forall n\in N, t\in T \tag{A4}$$

$$\beta_{k,n} = \frac{1}{|T|}\sum_{t\in T}\beta_{n,t}, \qquad \forall n\in N, k\in K_{individual} \tag{A5}$$

$$\beta_{k,t} = \frac{1}{|N|}\sum_{n\in N}\beta_{n,t}, \qquad \forall t\in T, k\in K_{trip} \tag{A6}$$

$$\beta_{k,0} = \frac{1}{|N||T|}\sum_{n\in N}\sum_{t\in T}\beta_{k,t}, \qquad \forall k\in K \tag{A7}$$

where $K$ is the set of parameters in the mode choice model, $\beta_0$ is a vector of $|K|$ population-level parameters, and $\beta_{n,t}$ is a vector of $|K|$ agent-level parameters. $K_{individual}$ is the set of parameters varying across individuals and $K_{trip}$ is the set of parameters varying across trips. Eq. (A1) defines the objective function, which is to minimize the variance of agent-level parameters. Eq. (A2) ensures that the observed choice $j^*$ maximizes the individual $n$'s utility in trip $t$, where $b$ is a safety margin with a recommended value around a predefined percentile of $(\varepsilon_{n,j^*,t}-\varepsilon_{n,j,t})$. Eqs. (A3) – (A4) determine the parameter boundaries for estimation, where $lb$ and $ub$ are vectors specifying the lower and upper bounds. Eq. (A5) ensures that individual-level parameters vary only across individuals. Eq. (A6) ensures that trip-level parameters vary only across trips. Eq. (A7) ensures that population-level (or generic) parameters are the average of agent-level parameters.

Solving the MIUM problem as a single QP problem would be computationally costly as it would lead to a highly sparse diagonal matrix. Instead, AMXL uses a decomposition method to initialize $\beta_0$ and update $\beta_0, \beta_{n,t}$ iteration by iteration. The subproblem with a fixed $\beta_0$ can be solved using any optimizer software or package that can handle QP like Gurobi, CVXPY, etc. The estimation algorithm is summarized in Algorithm 1.

**Algorithm 1.** Parameter estimation in the mode choice model

1. Given observed variables and mode choice, initialize with $i = 0$ and population-level parameters $\beta_0^{(i)} = [0,0,\dots,0]$.
2. For each trip $t \in T$ made by individual $n \in N$, solve a QP problem get $\beta_{n,t}^{(i)}$:
   $\min_{\beta_0,\beta_{n.t}} (\beta_0 - \beta_{n,t})^2$ subject to constraints in Eqs. (A2) – (A4).
3. Calculate individual-level parameters ($\beta_{k,n}$) and trip-level parameters ($\beta_{k,t}$) using Eqs. (A5) – (A6).
4. Set average to $y_0^{(i)} = \frac{1}{|N||T|}\sum_{n\in N}\sum_{t\in T}\beta_{k,t}$ as shown in Eq. (A7).
5. Use Method of Successive Averages (Sheffi & Powell, 1982) to update population-level parameters and get $\beta_0^{(i+1)}$:
   $$\beta_0^{(i+1)} = \frac{i}{i+1}\beta_0^{(i)} + \frac{1}{i+1}y_0^{(i)}$$
6. If the stopping criteria (percentage change smaller than 0.1%) for $\beta_0$ reached, stop and output $\beta_0^{(i)}, \beta_{n,t}^{(i)}$; else, set $i = i + 1$ and go back to Step 2.

Since only marginal subscription data is available at the upper branch, it is impossible to use maximum loglikelihood estimation (MLE) to estimate a choice model with confidence intervals. Instead, we calibrate the six subscription choice parameters ($f_{RP}$, $\beta_{gain}^{we}$, $\beta_{gain}^{wd}$, $\beta_{MT-user}$, $asc_{weekly}$, $asc_{monthly}$) by formulating a least squares optimization problem that minimizes the gap between predicted and observed number of subscribers. The optimization problem is shown in Eqs. (A8) – (A10).

$$\min_{\beta_{ridepass}} \sum_{j\in J_{ridepass}} \left(Num_j - \sum_{n\in N} P_{n,j}\right)^2 \tag{A8}$$

subject to:

$$P_{n,j} = \frac{e^{V_{n,j}}}{\sum_{j'\in J_{ridepass}} e^{V_{n,j'}}}, \qquad \forall n \in N, j \in J_{ridepass} \tag{A9}$$

$$V_{n,j} = \beta_{ridepass}^T X_{n,j}, \qquad \forall n \in N, j \in J_{ridepass} \tag{A10}$$

where $\beta_{ridepass}$ is a vector of six parameters serving as decision variables, $J_{ridepass}$ is the choice set that includes weekly, monthly, and none ride pass options, and $Num_j$ represents the observed number of individuals choosing a ride pass option. Eq. (A8) defines the objective function, which is to minimize the squared error of predicted subscribers. Eq. (A9) ensures that the probability $P_{n,j}$ is computed using the multinomial logit (MNL) form. Eq. (A10) ensures that the systematic utility $V_{n,j}$ equals the sum-product between vectors of parameters $\beta_{ridepass}$ and attributes $X_{n,j}$. We solve the optimization problem using the Generalized Reduced Gradient (GRG) nonlinear algorithm (Lasdon et al., 1974), which is one of the default methods in the solver tool in Microsoft Excel. Since our objective function is nonconvex, the GRG nonlinear algorithm has the risk of getting stuck at a local optimum. To reduce such an impact, we set bounds [-10, 10] for the ride pass parameters to avoid extreme values, and employ a multi-start routine. The results from different starting points did not change significantly in our experiments.

Estimating the lower and upper branches separately does not make sense since they are interrelated, i.e., subscription choice in the upper branch impacts microtransit trip fares, while mode choice in the lower branch provides information about gained benefits, microtransit user identity, and sensitivity to ride pass price. Hence, we propose an iterative algorithm to estimate them jointly. We start from assuming no individual holds a ride pass and setting initial parameters to zero. In each iteration, we estimate the lower-branch mode choice model with current microtransit trip fares, pass $Gain_n^{we}$, $Gain_n^{wd}$, $MT_n$, and $\beta_{price,n}$ to the upper-branch to estimate the subscription choice model, scale the lower-branch parameters using the nesting parameters, update the choice parameters using the Method of Successive Averages (MSA), and recalculate microtransit trip fares with the predicted ride pass ownership. This iterative process continues until the stopping criterion is satisfied. The joint estimation approach is summarized in Algorithm 2.

**Algorithm 2.** Joint estimation of subscription choice and mode choice

1. Initialize with no individual holding weekly or monthly ride pass ($P_{n,none} = 1, \forall n \in N$), calculate microtransit trip fares without a ride pass, and set $\beta_{ridepass}$, $\beta_{n,t}^{we}$, $\beta_{n,t}^{wd}$ to zero.
2. Estimate two lower-branch mode choice models (one for weekdays and the other for weekends) using Algorithm 1 and get parameters $\tilde{\beta}_{n,t}^{we}$, $\tilde{\beta}_{n,t}^{wd}$.
3. Calculate $Gain_n^{we}$, $Gain_n^{wd}$, $MT_n$, and $\beta_{price,n}$ based on Eqs. (4)-(6) in the main text.
4. Estimate the upper-branch subscription choice model by solving the problem in Eqs. (A8)-(A10) and get $\beta_{ridepass}$.
5. Scale mode choice parameters using nesting parameters:
   $\beta_{n,t}^{we} = \tilde{\beta}_{n,t}^{we}\beta_{gain}^{we}, \qquad \beta_{n,t}^{wd} = \tilde{\beta}_{n,t}^{wd}\beta_{gain}^{wd}, \qquad \forall n \in N, t \in T$
6. Update $\beta_{ridepass}$, $\beta_{n,t}^{we}$, $\beta_{n,t}^{wd}$ using Method of Successive Averages (MSA).
7. If the stopping criteria (percentage change smaller than some tolerance) for $\beta_{ridepass}$, $\beta_{n,t}^{we}$, $\beta_{n,t}^{wd}$ reached, stop and output these parameters; else recalculate microtransit trip fares with predicted ride pass ownership and go back to Step 2.

### A3. Replica data validation and visualization

Table A2 compares sociodemographic attributes of Replica's synthetic population with the 2024 ACS 5-Year Estimates obtained from the City of Arlington Open Data Portal (City of Arlington, 2026). Overall, the two datasets show strong agreement. Total population differs by only 0.59%. Employment rate and average household size also closely align, with the latter matching exactly at 2.74. Median household income differs by +4.07%, likely reflecting differences in inflation adjustment between the datasets. The homeownership rate shows the largest deviation among general attributes at +7.38%.

Moreover, the Replica and ACS data also show strong alignment in the proportions of these population segments. The not-low-income and low-income segments differ by +2.71% and +3.66%, while the senior and student segments differ by –4.85% and –5.30%. These slightly larger differences are expected because segment definitions combine multiple ACS variables (e.g., age, school enrollment, and poverty status).

**Table A2**
Comparison of sociodemographic attributes: Replica Data vs. ACS (2024)

| | Replica Data | ACS (2024) | % Difference |
|---|---|---|---|
| **Sociodemographic attributes** | | | |
| Total population | 400,103 | 397,742 | +0.59% |
| Female proportion | 51.01% | 50.71% | +0.59% |
| Employment rate | 65.83% | 66.72% | –1.33% |
| Median household annual income | $78,234 | $75,171 | +4.07% |
| Average household size | 2.74 | 2.74 | +0.00% |
| Rate of home ownership | 59.06% | 55.00% | +7.38% |
| **Proportion of population segments** | | | |
| Not-low-income segment | 54.52% | 53.08% | +2.71% |
| Low-income segment | 11.56% | 11.15% | +3.66% |
| Senior segment | 10.25% | 10.77% | –4.85% |
| Student segment | 23.67% | 24.99% | –5.30% |

Fig. A1 shows census block group-level OD pairs with more than 50 trips per day, illustrating diverse travel patterns on weekdays and weekends across the four population segments. Although trip survey data for Arlington are not available to directly validate Replica's synthetic trip data, the overall patterns align with expectations for a mid-sized city lacking fixed-route public transit — notably, the dominant driving mode share (~65%), the visibly denser and more spatially dispersed OD patterns for not-low-income segments compared to low-income and senior segments, and the concentrated travel patterns of students to the University of Texas at Arlington.

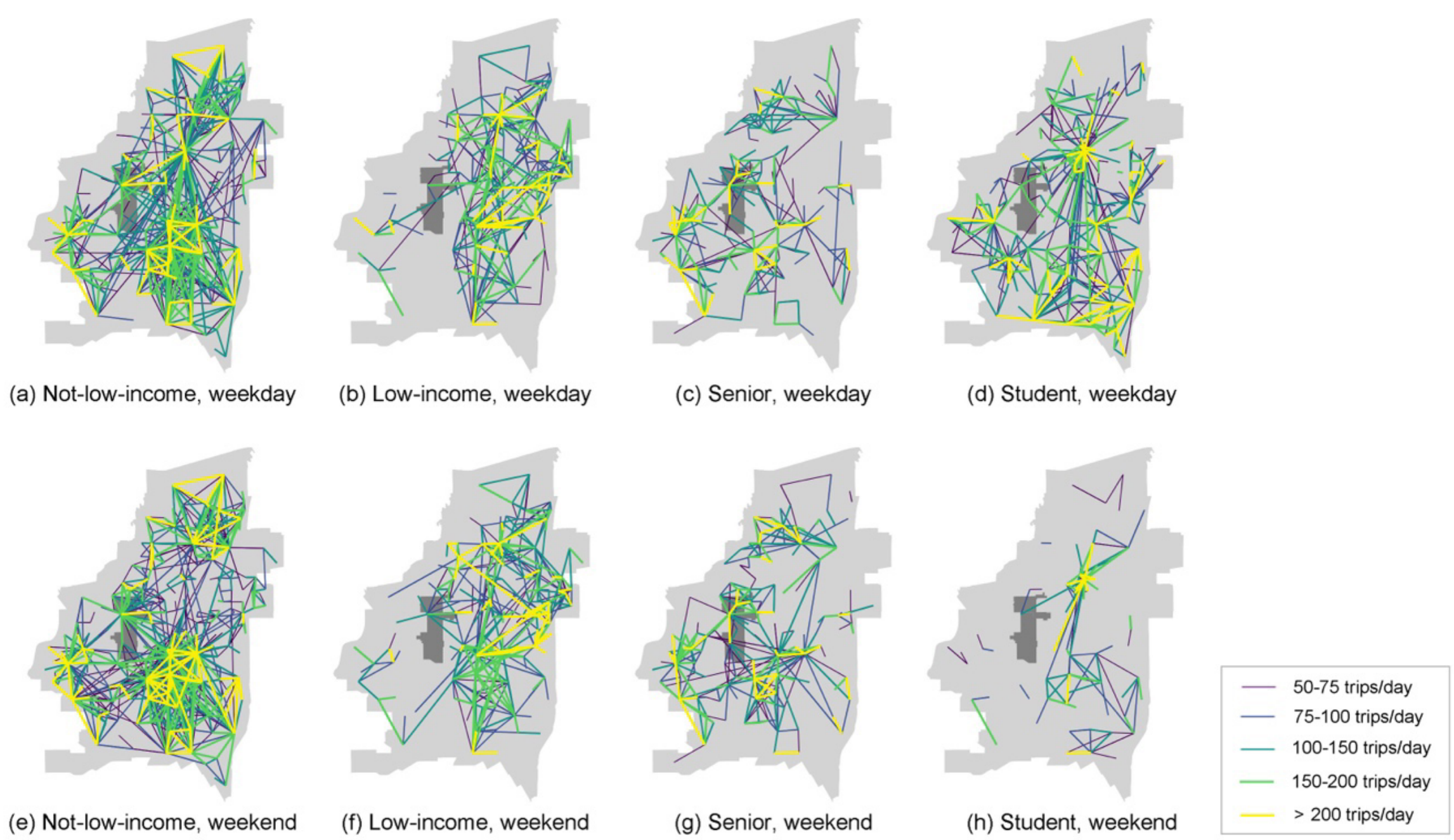

**Fig. A1.** Trips made by four population segments on weekdays and weekends.

### A4. Sigmoid structure and elasticity in the logit model

The multinomial logit model predicts the probability of choosing mode $i$ as:

$$P_i = exp(V_i)/\sum_{j \in J} exp(V_j) \tag{A9}$$

For a binary illustration, this reduces to the standard logistic (sigmoid) function $P = \frac{1}{1+e^{-x}}$, where $x$ represents the systematic utility difference. As shown in Fig. A2, the sigmoid curve has three distinct regions that directly explain the elasticity patterns observed in the sensitivity analysis (Section 4.2.4).

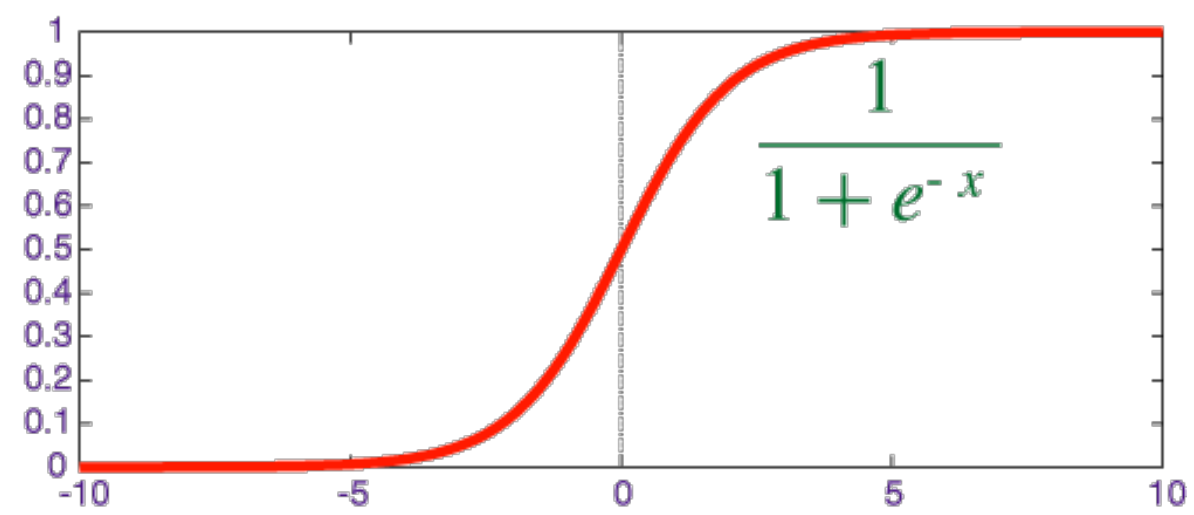

**Fig. A2.** Sigmoid function curve

In the left flat region (P ≈ 0), the choice probability is near zero and responds only marginally to changes in the utility argument. This corresponds to the current microtransit market share (<0.5%) in our case study. Because the predicted probability sits deep in this saturated tail, even substantial perturbations in in-vehicle time produce negligible shifts in mode share, ridership, and revenue—hence the near-zero elasticities observed for the "Current MT Market Share" scenario. In practical terms, the sigmoid acts as a natural buffer that dampens simulation errors rather than amplifying them.

In the steep middle region (P ≈ 0.5), the slope of the sigmoid is at or near its maximum, meaning that small changes in utility translate into large changes in predicted probability. This corresponds to the 20% market share scenario, which exhibits the highest elasticities across all three metrics. The mode's choice probability has moved off the flat tail and into the sensitive transition zone of the curve, where the model is most responsive to input perturbations.

In the right flat region (P ≈ 1), the curve begins to flatten again as the choice probability approaches its upper bound. This explains why the 40% market share scenario shows somewhat lower elasticities than the 20% case: the baseline probability has already entered the upper saturation region, where further utility improvements yield diminishing marginal gains in mode share. This non-monotonic pattern—low sensitivity at very low shares, peak sensitivity at moderate shares, and declining sensitivity at high shares—is an inherent property of the logistic function and generalizes to any multinomial logit application.